\documentclass[reqno,a4paper,11pt]{article}
\pdfoutput=1
\usepackage{xcolor}

\usepackage{graphicx}
\usepackage[textwidth = 430 pt, textheight = 630 pt]{geometry}

\definecolor{MyDarkBlue}{rgb}{0.15,0.25,0.45}
\usepackage{epsfig,rotating}
\usepackage{amsmath,amssymb}
\usepackage{amsfonts}
\usepackage{mathrsfs}
\usepackage{bbm}
\usepackage[normalem]{ulem}

\usepackage{latexsym}
\usepackage{amsthm}
\usepackage[all,knot]{xy}
\xyoption{arc}

\usepackage[utf8x]{inputenc}

\usepackage{hyperref}
\hypersetup{
hypertexnames=false,
colorlinks=true,
citecolor=MyDarkBlue,
linkcolor=MyDarkBlue,
urlcolor=MyDarkBlue,
pdfauthor={Christian S\"amann and Lennart Schmidt},
pdftitle={The Non-Abelian Self-Dual String},
pdfsubject={hep-th math-ph},
breaklinks=true
}

\usepackage{tikz}
\usepackage{mathtools}
\usepackage[all,knot]{xy}
\xyoption{arc}

\let\fn\footnote
\renewcommand{\footnote}[1]{\linespread{1.1}\fn{#1}\linespread{1.29}}

\makeatletter\renewcommand{\section}{\@startsection
{section}{1}{\z@}{-3.5ex plus -1ex minus
    -.2ex}{2.3ex plus .2ex}{\bf }}
\makeatletter\renewcommand{\subsection}{\@startsection{subsection}{2}{\z@}{-3.25ex
plus -1ex minus
   -.2ex}{1.5ex plus .2ex}{\bf }}
\makeatletter\renewcommand{\subsubsection}{\@startsection{subsubsection}{3}{-2.45ex}{-3.25ex
plus -1ex minus -.2ex}{1.5ex plus .2ex}{\it }}
\renewcommand{\thesection}{\arabic{section}}
\renewcommand{\thesubsection}{\arabic{section}.\arabic{subsection}}
\renewcommand{\@seccntformat}[1]{\@nameuse{the#1}.~~}

\renewcommand{\theequation}{\thesection.\arabic{equation}}
\makeatletter \@addtoreset{equation}{section}
\def\Ddots{\mathinner{\mkern1mu\raise\p@
\vbox{\kern7\p@\hbox{.}}\mkern2mu
\raise4\p@\hbox{.}\mkern2mu\raise7\p@\hbox{.}\mkern1mu}}
\usepackage[toc,page]{appendix}

\renewcommand{\thethm}{\thesection.\arabic{thm}}

\newcommand{\myxymatrix}[1]{\vcenter{\vbox{\xymatrix{#1}}}}

\renewcommand{\appendices}{
\section*{Appendix}\label{appendices}\setcounter{subsection}{0}
\addcontentsline{toc}{section}{Appendix}
\setcounter{equation}{0}
\makeatletter
\renewcommand{\theequation}{\Alph{subsection}.\arabic{equation}}
\renewcommand{\thesubsection}{\Alph{subsection}}
\renewcommand{\thethm}{\Alph{subsection}.\arabic{thm}}
\@addtoreset{equation}{subsection}
\@addtoreset{thm}{subsection}
\makeatother
}

\def\slasha#1{\setbox0=\hbox{$#1$}#1\hskip-\wd0\hbox to\wd0{\hss\sl/\/\hss}}

\def\periodb#1{\setbox0=\hbox{$#1$}#1\hskip-\wd0\hbox to\wd0{-}}
\newcommand{\nablas}{\slasha{\nabla}}
\newcommand{\dpars}{\slasha{\dpar}}

\newcommand{\unit}{\mathbbm{1}}   			
\newcommand{\id}{\mathrm{id}}   			

\newcommand{\CC}{\mathcal{C}}
\newcommand{\CCC}{\mathscr{C}}

\newcommand{\CF}{\mathcal{F}}

\newcommand{\CG}{\mathcal{G}}
\newcommand{\CCG}{\mathscr{G}}
\newcommand{\CH}{\mathcal{H}}

\newcommand{\CM}{\mathcal{M}}

\newcommand{\CN}{\mathcal{N}}

\newcommand{\CQ}{\mathcal{Q}}

\newcommand{\CR}{\mathcal{R}}

\newcommand{\CU}{\mathcal{U}}

\newcommand{\frg}{\mathfrak{g}}				

\newcommand{\FR}{\mathbbm{R}}     			
\newcommand{\NN}{\mathbbm{N}}     			
\newcommand{\RZ}{\mathbbm{Z}}     			

\newcommand{\lambdab}{\bar{\lambda}}

\newcommand{\dd}{\mathrm{d}}     			
\newcommand{\dpar}{\partial}     			
\newcommand{\embd}{{\hookrightarrow}}     		
\newcommand{\di}{\mathrm{i}}     			
\newcommand{\eps}{{\varepsilon}}			

\newcommand{\sB}{\mathsf{B}}

\newcommand{\etab}{{\bar{\eta}}}

\newcommand{\eor}{{\qquad\mbox{or}\qquad}}     		
\newcommand{\eand}{{\qquad\mbox{and}\qquad}}     		
\newcommand{\ewith}{{\qquad\mbox{with}\qquad}}
\newcommand{\efor}{{\qquad\mbox{for}\qquad}}

\newcommand{\der}[1]{\frac{\dpar}{\dpar #1}}   		
\newcommand{\dder}[1]{\frac{\dd}{\dd #1}}   		
\newcommand{\tr}{\,\mathrm{tr}\,}     			
\newcommand{\pr}{\mathsf{pr}}     			

\newcommand{\au}{\mathfrak{u}}
\newcommand{\asu}{\mathfrak{su}}

\newcommand{\fra}{\mathfrak{a}}
\newcommand{\aspin}{\mathfrak{spin}}
\newcommand{\astring}{\mathfrak{string}}

\newcommand{\sU}{\mathsf{U}}     			

\newcommand{\sG}{\mathsf{G}}

\newcommand{\sL}{\mathsf{L}}

\newcommand{\sSU}{\mathsf{SU}}

\newcommand{\sO}{\mathsf{O}}
\newcommand{\sE}{\mathsf{E}}

\newcommand{\CatLietwo}{\mathsf{Lie2alg}}
\newcommand{\CatMfd}{\mathsf{Mfd}^\infty}

\newcommand{\sSO}{\mathsf{SO}}
\newcommand{\sSpin}{\mathsf{Spin}}
\newcommand{\sSp}{\mathsf{Sp}}
\newcommand{\sString}{\mathsf{String}}

\newcommand{\comment}[1]{}     				
\def\tyng(#1){\hbox{\tiny$\yng(#1)$}}			
\def\tyoung(#1){\hbox{\tiny$\young(#1)$}}			

\newcommand{\beq}{\begin{eqnarray}}
\newcommand{\eeq}{\end{eqnarray}}

\newcommand{\sft}{{\sf t}}
\newcommand{\sfg}{{\sf g}}
\newcommand{\sfh}{{\sf h}}
\newcommand{\sfd}{{\sf d}}
\newcommand{\sfb}{{\sf b}}
\newcommand{\sff}{{\sf f}}

\newcommand{\sfs}{\mathsf{s}}

\definecolor{outrageousorange}{rgb}{1.0, 0.43, 0.29}

\newcommand{\CatGrp}{\mathsf{Grp}}

\begin{document}
\begin{titlepage}
\begin{flushright}
 EMPG--17--05
\end{flushright}
\vskip2.0cm
\begin{center}
{\LARGE \bf The Non-Abelian Self-Dual String}
\vskip1.5cm
{\Large Christian S\"amann and Lennart Schmidt}
\setcounter{footnote}{0}
\renewcommand{\thefootnote}{\arabic{thefootnote}}
\vskip1cm
{\em Maxwell Institute for Mathematical Sciences\\
Department of Mathematics, Heriot--Watt University\\
Colin Maclaurin Building, Riccarton, Edinburgh EH14 4AS, U.K.}\\[0.5cm]
{Email: {\ttfamily c.saemann@hw.ac.uk~,~ls27@hw.ac.uk}}
\end{center}
\vskip1.0cm
\begin{center}
{\bf Abstract}
\end{center}
\begin{quote}
We argue that the relevant higher gauge group for the non-abelian generalization of the self-dual string equation is the string 2-group. We then derive the corresponding equations of motion and discuss their properties. The underlying geometric picture is a string structure, i.e.~a categorified principal bundle with connection whose structure 2-group is the string 2-group. We readily write down the explicit elementary solution to our equations, which is the categorified analogue of the 't Hooft--Polyakov monopole. Our solution passes all the relevant consistency checks; in particular, it is globally defined on $\FR^4$ and approaches the abelian self-dual string of charge one at infinity.  We note that our equations also arise as the BPS equations in a recently proposed six-dimensional superconformal field theory and we show that with our choice of higher gauge structure, the action of this theory can be reduced to four-dimensional supersymmetric Yang--Mills theory.
\end{quote}
\end{titlepage}

\tableofcontents

\newpage

\section{Introduction and discussion of results}

In this paper, we derive the appropriate higher analogue of the Bogomolny monopole equation and give the elementary solution. Our guidelines come from string theory as well as the framework called {\em higher gauge theory}~\cite{Baez:2002jn,Baez:2004in,Baez:0511710,Sati:2008eg,Baez:2010ya,Saemann:2016sis}. In this setting, connections on principal fiber bundles with Lie structure groups are replaced by categorified connections on non-abelian variants of gerbes with structure Lie 2-groups. 

\subsection{Motivation}

The motivation for our work is (at least) threefold: First, recall that BPS monopoles can be described in string theory by D1-branes ending on D3-branes. A lift to M-theory leads to configurations called {\em self-dual strings}, which are given by M2-branes ending on M5-branes~\cite{Howe:1997ue}. These self-dual strings should be BPS states in a long-sought six-dimensional $\CN=(2,0)$ superconformal field theory, often simply referred to as the {\em (2,0)-theory}. This still rather poorly understood theory is of great importance because it should provide an effective description of stacks of multiple M5-branes, just as maximally supersymmetric Yang--Mills theory does for D-branes. A classical description of its BPS states would certainly advance our understanding.

A second motivation stems from the development and study of higher integrable models. The BPS monopole equation is an example of a classical integrable system, just as the self-dual Yang--Mills or instanton equation in four dimensions as well as Hitchin's vortex equations. This means that it has rich underlying geometric structures that allow for a relatively explicit description of the solutions and their moduli space. Among these geometric structures are twistor descriptions as well as the {\em Nahm transform} which, in an extreme variant, generates solutions to the Bogomolny monopole equation from solutions to a one-dimensional equation via zero-modes of a Dirac operator. Higher and non-abelian generalizations of integrable systems exist, and their moduli spaces have been described using twistor methods~\cite{Saemann:2012uq,Saemann:2013pca,Jurco:2014mva,Jurco:2016qwv}. The corresponding higher Nahm transform would certainly be very interesting in its own right and yield further insights into the dualities of M-theory. Most interestingly for mathematicians is that it would provide us with a natural candidate for a categorified Dirac operator, a very important and still missing ingredient in elliptic cohomology. 

The third motivation comes from higher differential geometry. Abelian gerbes have become an important tool in areas such as twisted K-theory and many interesting examples are known, often of relevance in string theory. The situation is very different for non-abelian gerbes: we are not aware of any other non-trivial and truly non-abelian gerbe with connection of relevance to string or M-theory beyond the one presented in this paper. Without explicit examples, however, it is difficult to develop a mathematical area to its full potential, and this seems to have been a problem of higher gauge theory in the past. This is particularly true since there is a widespread belief that non-abelian gerbes are essentially abelian and therefore cannot contribute new aspects to mathematical physics.

\subsection{Results}

We start by considering the principal fiber bundles underlying the abelian Dirac monopole and the non-abelian 't Hooft--Polyakov monopole. We observe that the appropriate gauge group for the 't Hooft--Polyakov monopole is the total space $S^3\cong \sSU(2)$ of the principal bundle describing the Dirac monopole. In particular, there is a bundle morphism embedding the latter in the trivial $\sSU(2)$-bundle which preserves all topological information. We then lift this picture to the self-dual string, which suggests that the appropriate higher gauge group or gauge 2-group should be the total space $\CCG_{\rm F}$ of the fundamental gerbe over $S^3$, which can be endowed with a higher group structure. The result is a 2-group model of the string group $\sString(3)$, which is also a very natural candidate for a gauge structure from a number of different perspectives. 

Working with higher or categorified structures implies working with larger classes of equivalences. In particular, the 2-group model of $\sString(3)$ can be cast in several equivalent, but quite different looking forms. To demonstrate that our constructions are reasonable, we will choose to work with two extreme models in parallel: the strict 2-group model $\sString_{\hat\Omega}(3)$ employing path and loop spaces~\cite{Baez:2005sn} as well as the semistrict and finite dimensional model $\sString_{\rm sk}(3)$ of~\cite{Schommer-Pries:0911.2483}. Since our aim is a self-dual string equation on the contractible space $\FR^4$, all objects are local and we can directly switch to the corresponding Lie 2-algebra models $\astring_{\hat\Omega}(3)$ and $\astring_{\rm sk}(3)$.

Given an arbitrary higher gauge algebra, one can straightforwardly derive the corresponding local notions of gauge potentials, curvatures, gauge transformations, Bianchi identities and topological invariants. For the string Lie 2-algebra models, we find that the result is not suitable for a description of non-abelian self-dual strings. Instead, one should modify the definitions of the curvatures as done in the case of twisted string structures~\cite{Sati:2009ic}. We shall indicate the twist by a superscript $T$ attached to the gauge Lie 2-algebra. In the case of the finite-dimensional Lie 2-algebra model $\astring^T_{\rm sk}(3)$, the gauge field content consists of an ordinary, $\asu(2)$-valued gauge potential $A$ and an abelian $B$-field. The curvatures read as
\begin{equation}\label{eq:sds_eom_intro}
 F=\dd A+\tfrac12[A,A]~,~~~H=\dd B+\tfrac{1}{3!}(A,[A,A])-(A,F)=\dd B-(A,\dd A)-\tfrac13(A,[A,A])~,
\end{equation}
where $(-,-)$ is the Killing form on $\asu(2)$ and the additional term $(A,F)$ arises when twisting the ordinary higher curvature. The corresponding Bianchi identity reads as $\dd H=(F,F)$, which suggests the following non-abelian self-dual string equations:
\begin{equation}
 H=*\dd \varphi\eand F=*F~,
\end{equation}
where $\varphi$ is an abelian Higgs field. These equations are indeed gauge invariant. They clearly contain the abelian self-dual string and they are compatible with string theory expectations. As we find, they nicely reduce to the monopole equations in three dimensions and they appear as the BPS equations for a suitable choice of gauge structure in the superconformal (1,0)-model of~\cite{Samtleben:2011fj}, see also~\cite{Akyol:2012cq}. Closely related configurations were considered very recently in a different model~\cite{Singh:2017viv}, where they were called I-strings. All these constructions can now be properly understood from the perspective of (twisted) higher gauge theory. Finally, it is encouraging to see a Chern--Simons term appear in the 3-form curvature. This points towards a close link to the M2-brane models~\cite{Bagger:2007jr,Gustavsson:2007vu,Aharony:2008ug}, which are Chern--Simons matter theories.

A similar modification of the 3-form curvature as in~\eqref{eq:sds_eom_intro} leads to analogous results in the case of the loop space model $\astring_{\hat\Omega}(3)$ and there is a one-to-one correspondence between gauge equivalence classes of solutions to both equations arising from the expected categorical equivalence. 

Interestingly, the modifications of the 3-form curvatures for both string Lie 2-algebra models $\astring_{\rm sk}(3)$ and $\astring_{\hat \Omega}(3)$ change the infinitesimal gauge transformations of $H$ from\footnote{cf.~equation~\eqref{eq:naive_gauge_trafos_curvatures}}
\begin{equation}
 \delta H =\mu_2(H,\alpha)+\mu_2(\CF,\Lambda)-\mu_3(\CF,A,\alpha)
\end{equation}
to $\delta H=0$. This fixes the broken gauge invariance of one of the first non-abelian higher gauge theories written down~\cite{Baez:2002jn}. It also renders the non-abelian self-duality equation $H=*H$ in six dimensions gauge invariant beyond the case $\CF=0$.

We now readily write down the charge one solution to~\eqref{eq:sds_eom_intro} based on the elementary $\sSU(2)$-instanton. The resulting fields are all non-singular over $\FR^4$ and interacting in the sense that non-trivial linear combinations of this solution are no longer solutions. This clearly shows that our result is not an abelian solution simply recast in an unusual form, but rather a genuinely non-abelian self-dual string. At infinity, the solution approaches the abelian self-dual string and the obvious considerations of topological charges work as expected. As a last consistency check, we note that categorical equivalence between $\astring_{\rm sk}(3)$ and $\astring_{\hat\Omega}(3)$ can be used to map this solution to a solution of the self-dual string equations for $\astring^T_{\hat\Omega}(3)$.

While our discussion on $\FR^4$ is in principle consistent, the instanton solution suggests that we are actually working over $S^4$. There, the first fractional Pontryagin class $\tfrac12 p_1=(F,F)$ is clearly not trivial in $H^4(S^4,\RZ)$. Mathematical string structures (i.e.~principal 2-bundles with structure 2-group the string 2-group) as defined in~\cite{Killingback:1986rd,Stolz:2004aa,Redden:2006aa,Waldorf:2009uf,Sati:2009ic}, however, require $[\tfrac12 p_1]=0$. In particular, a string structure is encoded in a spin structure and a trivialization of $\tfrac12 p_1$. 

We note that this issue can be solved by extending our structure Lie 2-algebra from $\astring_{\rm sk}^T(3)$ to $\astring_{\rm sk}^T(4)$, where the latter has underlying Lie algebra $\aspin(4)\cong \asu(2)\times \asu(2)$. In particular, we extend our equations~\eqref{eq:sds_eom_intro} to
\begin{equation}
 F_L=*F_L~,~~~F_R=-*F_R~,~~~[\tfrac12 p_1]=[(F,F)]=0~,~~~H=*\dd \varphi~.
\end{equation}
Here, $F_L$ and $F_R$ are the components of the 2-form curvatures taking values in the two $\asu(2)$-factors of $\aspin(4)$ and their total should have vanishing first Pontryagin class. The field $H$ is then globally defined, and yields a trivialization of $\tfrac12 p_1=(F,F)=-\dd H$.

We can directly extend our previous solution to an explicit string structure as follows. Let $F_L$ and $F_R$ be the curvatures of a charge one instanton and a charge one anti-instanton, respectively, which can be centered at different locations. Then $[\tfrac12 p_1]=1-1=0$ and the expressions for the trivialization $H$ and the scalar field $\varphi$ are readily computed. In particular, $H=0$ and $\varphi=0$ if instanton and anti-instanton are of the same size and centered at the same point.

Since our construction of the self-dual string fulfills our physical and mathematical expectations, we can move on and ask what we can learn about the full (2,0)-theory. Here, the most evident candidate for further progress is the (1,0)-model\footnote{Considering $(1,0)$-theories is indeed very natural: First, $(2,0)$-theories are special cases of $(1,0)$-theories. Second, recall that general M2-brane models were expected to posses $\CN=8$ supersymmetry in three dimensions. The ABJM model, however, only features $\CN=6$ supersymmetry, and full supersymmetry is restored by non-local ``monopole operators.'' The six-dimensional analogue would be a $(1,0)$-theory whose supersymmetry is enhanced to that of the $(2,0)$-theory by non-local ``self-dual string operators.''} of~\cite{Samtleben:2011fj}, which was derived from the closure of non-abelian generalizations of supersymmetry transformations of supergravity tensor hierarchies.\footnote{A closely related model was derived in~\cite{Chu:2011fd}.} It is already known that this model is based on an underlying higher Lie algebra which is endowed with some additional structure maps~\cite{Palmer:2013pka,Lavau:2014iva}; also the relations to ordinary higher gauge theory were pointed out~\cite{Palmer:2013pka}. To use the Lie 2-algebra $\astring^T_{\rm sk}(3)$ in the (1,0)-model, we require an inner product structure on it. The appropriate notion of {\em cyclic inner product} is most naturally encoded in a symplectic structure on the grade-shifted $L_\infty$-algebra, and we therefore need to double $\astring^T_{\rm sk}(3)$ in an evident way to a Lie 3-algebra $\widehat{\astring}_\omega(3)$. We find that the result is indeed a valid gauge structure for the (1,0)-model. Note that it has been observed before that the string Lie 2-algebra is a suitable (1,0) gauge structure~\cite{Akyol:2012cq,Palmer:2013pka}. However, it seems that the relevance of this Lie 2-algebra has been underappreciated due to the resulting 3-form curvature taking values in $\FR\cong \au(1)$.

Our observations clarify at least two important problems with the (1,0)-model. First, it was not clear what the appropriate gauge structure for the (1,0)-model was supposed to be. The string Lie 2-algebra, as well as its twist and doubling to a Lie 3-algebra, exists for any Lie algebra of ADE type. Therefore, the doubling yields natural candidates for higher gauge structures for all the ADE-classified $(2,0)$-theories. Second, the (1,0)-model had an instability due to a cubic scalar field interaction term. This term vanishes in the case of the doubled string Lie 2-algebra.

An important consistency check that no classical candidate for a non-abelian $(2,0)$-theory has passed so far is a consistent reduction process to supersymmetric Yang--Mills theory. We find that the (1,0)-model based on the doubled string Lie 2-algebra $\widehat{\astring}_\omega(3)$ reduces to $\CN=2$ super Yang--Mills theory in four dimensions by a strong coupling expansion similar in spirit to the reduction of M2-brane models to super Yang--Mills theory~\cite{Mukhi:2008ux}. While this reduction still lacks features of the expected reduction of the (2,0)-theory to Yang--Mills theory, it is certainly a step forward.

\subsection{Outlook}

In this paper, we have established the classical existence of truly non-abelian self-dual strings. This means that there are interesting, non-trivial and physically relevant higher gauge theories. It also implies that one might be able to learn more about the six-dimensional $\CN=(2,0)$ superconformal field theory than commonly thought. In particular, a classical description of the dynamics of multiple coinciding M5-branes might become similarly feasible as in the case of coinciding M2-branes by introducing categorified differential geometric notions.

Our results lead directly to a number of concrete and obvious open questions that should be studied in detail. First, it is certainly worthwhile to explore the relation to the ABJM model in detail. Since the latter model is a higher gauge theory~\cite{Palmer:2013ena} and contains a Chern--Simons term, there is now hope that a link to boundary ABJM theory analogous to e.g.~\cite{Chu:2009ms} is possible. Second, one should consider more solutions and try to capture them by modifying the twistor constructions of~\cite{Saemann:2012uq,Saemann:2013pca,Jurco:2014mva,Jurco:2016qwv}. We note that the twisted string structure also yields a gauge invariant self-duality equation in six dimensions, whose solutions should be of interest, too. Third, the explicit form of the elementary self-dual string is now known explicitly, which opens up the road to finding the appropriate higher version of the Nahm transform. Fourth, our result sheds some light on the quantization of 2-plectic manifolds: The corresponding boundary M2-brane model should describe the proper higher quantization of a 3-sphere, along similar lines as those of~\cite{Bunk:2016rta}. The answers to this issue available in the literature do not seem to be satisfying to us. Fifth, one should use our example of a (1,0) gauge structure to revisit the (1,0)-model of~\cite{Samtleben:2011fj} and to try to resolve its remaining issues.

\subsection{Reading guide}

While many of our constructions involve mathematical notions that may not be familiar to theoretical physicists, the results should be relatively easy to understand. Here are a few points to help navigate the paper without any understanding of higher or categorified geometry and algebra:
\begin{itemize}
 \setlength{\itemsep}{-0.5mm}
 \item[$\triangleright$] A first strong reason for choosing our gauge structure, the string Lie 2-group, is given in section~\ref{ssec:geometric_picture}; further arguments that this is the appropriate one are summarized in section~\ref{ssec:arguments_for_string}. If the reader is happy with our choice of gauge structure, section~\ref{sec:SDS_geometry} can be skipped.
 \item[$\triangleright$] Our untwisted and twisted gauge structure exists in two different, but equivalent formulations labeled $\astring_{\rm sk}(3)$ and $\astring_{\hat\Omega}(3)$ as well as $\astring^T_{\rm sk}(3)$ and $\astring^T_{\hat\Omega}(3)$. While the fact that our discussion always preserves categorical equivalence between these pairs is an important consistency check, we recommend to focus on the case $\astring^T_{\rm sk}(3)$ at first.
 \item[$\triangleright$] A very brief introduction to the kinematical data of higher gauge theory with a derivation of gauge potentials, curvatures and gauge transformations is found in section~\ref{ssec:higher_gauge_theory}. An equivalent, but more modern and more powerful approach is outlined in section~\ref{ssec:hgt_dgas}. 
 \item[$\triangleright$] The canonical structures obtained as such require, however, a further twist to be suitable for a description of non-abelian self-dual strings. For $\astring_{\rm sk}(3)$, this twist is discussed in section~\ref{ssec:twist}, while for $\astring_{\hat\Omega}(3)$, the twist is found directly in section~\ref{ssec:sds_eom2}. One can skip the motivation for the twist and directly go to the equations of motion presented in sections~\ref{ssec:sds_eom} and~\ref{ssec:sds_eom2}.
 \item[$\triangleright$] The explicit elementary solution for $\astring^T_{\rm sk}(3)$ is given in section~\ref{ssec:solution}, where it is also shown that this solution has the expected properties.
  \item[$\triangleright$] The global geometric picture is discussed in section~\ref{sec:global_picture} in a way that requires little knowledge of 2-gerbes.
 \item[$\triangleright$] The fact that our gauge structure readily extends to a suitable gauge structure for the (1,0)-model obtained from tensor hierarchies is explained in section~\ref{sec:10model}. While the extension presented in section~\ref{ssec:ext_Lie_3} requires some knowledge of N$Q$-manifolds, the rest of the section including the reduction to four-dimensional super Yang--Mills theory should be comprehensible without this.
 \item[$\triangleright$] N$Q$-manifolds are introduced merely as a technical tool, anything in the discussion related to them can safely be ignored. Even for many of our arguments, knowledge of $L_\infty$-algebras is sufficient.
 \item[$\triangleright$] The appendix gives a brief introduction to higher or categorified algebra and geometry, which makes the paper almost self-contained. Also, a number of hopefully helpful references to more detailed explanations of the required background material are provided. 
\end{itemize}

\section{The self-dual string and the string Lie 2-algebra}\label{sec:SDS_geometry}

In this section, we develop the topological picture of a potential non-abelian self-dual string solution, drawing on analogies with monopoles.

\subsection{Geometric description with principal 2-bundles}\label{ssec:geometric_picture}

Recall that the Dirac monopole is described by the principal $\sU(1)$-bundle $P_{\rm D}$ corresponding to the Hopf fibration:
\begin{equation}\label{eq:geom_Dirac}
\myxymatrix{
 \sU(1) \ar@{^{(}->}[r] & S^3\cong \sSU(2)\ar@{->}[d]^\pi \\
 & S^2 \cong \frac{\sSU(2)}{\sU(1)}}
\end{equation}
This is the principal bundle over $S^2$ with first Chern class~1. Because principal $\sU(1)$-bundles over $S^2$ are characterized by $H^2(S^2,\RZ)\cong \RZ$, this bundle, together with its dual, generates all possible principal $\sU(1)$-bundles over $S^2$ via tensor products.

Since this bundle is non-trivial, it cannot be extended from $S^2\embd \FR^3$ to all of $\FR^3$, but only to $\FR^3\backslash\{x\}$ for some $x\in\FR^3$.\footnote{Inversely since $\FR^3\backslash\{x\}$ is homotopy equivalent to $S^2$, any principal bundle on $\FR^3\backslash\{x\}$ originates, up to isomorphism, from a pullback along the embedding $S^2\embd \FR^3\backslash\{x\}$. This extends in an obvious way to $\FR^{n+1}\backslash\{x\}$ and $S^n$ as well as to higher principal bundles.} This fact is reflected in the singularity of the gauge field description of the Dirac monopole in which the Higgs field has a singularity at the point $x$. 

The 't Hooft--Polyakov monopole~\cite{Hooft:1974:276,Polyakov:1974ek,Prasad:1975kr}, on the other hand, can be extended from $S^2$ to all of $\FR^3$, since it corresponds to a non-trivial field configuration whose underlying geometry is the trivial principal $\sSU(2)$-bundle $P_{\rm HP}$. We can make the transition from the Dirac monopole to the 't Hooft--Polyakov monopole preserving all topological information by embedding $P_{\rm D}$ into $P_{\rm HP}$ via a homomorphism of principal bundles. Recall that such a morphism $P_{\rm D}\embd P_{\rm HP}$ involves a homomorphism of Lie groups $\rho:\sU(1)\rightarrow \sSU(2)$. The remaining maps are read off the following diagram:
\begin{equation}
\myxymatrix{
 \sU(1) \ar@{^{(}->}[r]\ar@{-->}@/^5ex/[rrrr]^{\rho} & \sSU(2)\ar@{->}[d]^\pi \ar@{-->}@/^5ex/[rrrr]^{\pi\times\id} & & & \sSU(2) \ar@{^{(}->}[r] & S^2\times \sSU(2)\ar@{->}[d]^{\rm pr}\\
 & S^2\ar@{-->}@/^3ex/[rrrr]^{\id} & & & & S^2}
\end{equation}
It is well-known that this embedding even describes the underlying gauge field description of the monopole asymptotically.

Let us now try to find a corresponding picture for the self-dual string. The abelian self-dual string soliton involves a 3-form curvature which is the curvature of the connective structure of an abelian gerbe or principal 2-bundle over $S^3$.

Principal 2-bundles are defined as categorifications of ordinary principal bundles. That is, we start from the notion of a 2-space\footnote{A slight generalization of a Lie groupoid: A 2-space is a category whose objects and morphisms form smooth manifolds and whose source, target, identity and composition maps are smooth. A simple notion of a map between 2-spaces is given by a functor consisting of smooth maps.} and use it to define Lie 2-groups and fibrations of 2-spaces. The latter give rise to 2-actions and 2-covers, and we can follow the usual definition of a principal bundle as done in~\cite{Wockel:2008aa} or in~\cite{Waldorf:1608.00401} for connections. 

An important example of a Lie 2-group is $\sB\sU(1)=(\sU(1)\rightrightarrows *)$ and principal $\sB\sU(1)$-bundles are often called abelian gerbes. Let $Y\twoheadrightarrow M$ be a cover of a manifold $M$ with $Y=\sqcup_a U_a$ the disjoint union of the patches $U_a$. Define $Y^{[n]}\coloneqq Y\times_M Y\times_M\dots \times_M Y$ to be the $n$-fold fiber product or, equivalently, the disjoint union of the intersections of any $n$ patches. Recall that a principal $\sU(1)$-bundle corresponds to a bundle over $Y$, whose components are glued together by bundle isomorphisms over $Y^{[2]}$, which satisfy a cocycle condition over $Y^{[3]}$. Analogously, a principal $\sB\sU(1)$-bundles is given by bundles over $Y^{[2]}$, glued together by bundle isomorphisms over $Y^{[3]}$, which satisfy a cocycle condition over $Y^{[4]}$. While principal $\sU(1)$-bundles are characterized by elements in $H^2(M,\RZ)$, principal $\sB\sU(1)$-bundles are characterized by elements in $H^3(M,\RZ)$ called the {\em Dixmier--Douady class} of the abelian gerbe. More details and references are given in appendix~\ref{app:A}.

The self-dual string now corresponds to a categorified Dirac monopole over $S^3$, and we have the following analogue of~\eqref{eq:geom_Dirac} in terms of 2-spaces and maps between these:
\begin{equation}
\myxymatrix{
 \sB\sU(1)\ar@{^{(}->}[r] & \CCG_{\rm F}\ar@{->}[d]^\pi \\
 & (S^3\rightrightarrows S^3) }
\end{equation}
Here, we regarded the space $S^3$ as the trivial 2-space $S^3\rightrightarrows S^3$. The abelian gerbe $\CCG_{\rm F}$ has Dixmier--Douady class~1. Since $H^3(S^3,\RZ)\cong \RZ$, this principal $\sB\sU(1)$-bundle, together with its dual, generates all principal $\sB\sU(1)$-bundles over $S^3$. Again, because $\CCG_{\rm F}$ is topologically non-trivial, it cannot be extended from $S^3\embd \FR^4$ to all of $\FR^4$, but just to $\FR^4\backslash\{x\}$ with $x\in \FR^4$. Just as in the case of the Dirac monopole, this is reflected in a singularity of the Higgs field at $x$ in the corresponding field description.

To follow the analogy with the 't Hooft--Polyakov monopole, we observe that the manifold underlying the relevant gauge group of the non-abelian monopole is simply the total space of $P_{\rm D}$. Correspondingly, we wish to embed $\CCG_{\rm F}$ into a non-abelian principal 2-bundle with a structure 2-group which has as its underlying 2-space $\CCG_{\rm F}$. Together with a homomorphism of Lie 2-groups $\rho:\sB\sU(1)\rightarrow \CCG_{\rm F}$, this embedding is then given by
\begin{equation}
\myxymatrix{
 \sB\sU(1) \ar@{^{(}->}[r]\ar@{-->}@/^5ex/[rrrr]^{\rho} & \CCG_{\rm F}\ar@{->}[d]^\pi \ar@{-->}@/^5ex/[rrrr]^{\pi\times\id} & & & \CCG_{\rm F} \ar@{^{(}->}[r] & (S^3\rightrightarrows S^3)\times \CCG_{\rm F}\ar@{->}[d]^{\rm pr}\\
 & (S^3\rightrightarrows S^3)\ar@{-->}@/^3ex/[rrrr]^{\id} & & & & (S^3\rightrightarrows S^3)}
\end{equation}
Note that all the maps in the above picture are now homomorphisms of 2-spaces.

For this picture to make sense, we need a Lie 2-group structure on the 2-space $\CCG_{\rm F}$. This structure exists, and it completes $\CCG_F$ to what is known as a 2-group model of $\sString(3)$.

\subsection{String 2-group models}

The string group $\sString(n)$ fits into the following sequence, known as the Whitehead tower of $\sO(n)$:
\begin{equation}
 \dots \rightarrow \sString(n)\rightarrow \sSpin(n)\rightarrow \sSpin(n)\rightarrow \sSO(n)\rightarrow \sO(n)~.
\end{equation}
The arrows describe isomorphisms at the level of homotopy groups, with the exception of the lowest homotopy group: $\pi_0(\sO(n))$ is removed in the step from $\sO(n)$ to $\sSO(n)$, $\pi_1(\sO(n))$ in the step to $\sSpin(n)$ and $\pi_2(\sO(n))$ is trivial, anyway. The string group $\sString(n)$ is obtained by removing $\pi_3(\sO(n))$. That is, $\sString(n)$ is a 3-connected cover of $\sSpin(n)$. This definition determines $\sString(n)$ only up to homotopy, and the string group structure is therefore only determined up to $A_\infty$-equivalence.

Note that $\sString(n)$ cannot be a finite-dimensional Lie group, since $\pi_1$ and $\pi_3$ of $\sString(n)$ are trivial. First models of the string group were presented in~\cite{Stolz:1996:785-800,Stolz:2004aa}. These models are slightly inconvenient since the underlying space is only determined up to a weak homotopy equivalence and only the homotopy type of the fiber $\sString(n)\rightarrow \sSpin(n)$ is determined. This situation is ameliorated when considering 2-group models of the string group, which are Lie 2-groups endowed with a Lie 2-group homomorphism to $\sSpin(n)$. Besides the better control over the fibers and the gained natural smooth structure, there are a number of further reasons for considering Lie 2-group models. First, we can choose to work with a finite dimensional Lie 2-group model, leading to a corresponding finite dimensional Lie 2-algebra, which turns out to be convenient for certain purposes. More importantly, however, the 2-group refinement is very natural from the string theory perspective, where the expected higher gauge theory requires a higher gauge group. Also, this refinement of the string group is induced by the second Chern class and therefore tightly linked to instantons, which are expected to feature in the non-abelian self-dual string.

There are various such Lie 2-group models but for our considerations, those of~\cite{Schommer-Pries:0911.2483} and~\cite{Baez:2005sn} will suffice. We will refer to these as the {\em skeletal model} $\sString_{\rm sk}(n)$ and the {\em loop space model} $\sString_{\hat \Omega}(n)$, respectively.

The two string 2-group models of $\sString(3)$ start from two different but equivalent ways of describing the fundamental bundle gerbe $\CCG_{\rm F}$ over $S^3\cong \sSU(2)\cong \sSpin(3)$. The skeletal model uses a suitable\footnote{A cover of $\sSpin(3)$ that extends to a simplicial cover of the nerve of $\sB\sSpin(3)=(\sSpin(3)\rightrightarrows *)$.}, ordinary cover $Y_1=\CU=\sqcup_a U_a$ of $\sSU(2)$, while the loop space model starts from the surjective submersion $Y_2=P_0\sSU(2)\stackrel{\dpar}{\twoheadrightarrow} \sSU(2)$, where
\begin{equation}
 P_0\sSU(2)\coloneqq \{\gamma:[0,1]\rightarrow \sSU(2)~|~\gamma(0)=\unit\}~,~~~\dpar\gamma\coloneqq \gamma(1)~.
\end{equation}
Note that $Y_2^{[2]}$ is the space of pairs of paths with the same endpoint. Modulo smoothness at the endpoints, which is a technicality we shall suppress, this yields the space of based, parameterized loops,
\begin{equation}
 \Omega\sSU(2)\coloneqq \{\ell:[0,1]\rightarrow \sSU(2)~|~\ell(0)=\ell(1)=\unit\}~.
\end{equation}
We then have the following two descriptions of $\CCG_{\rm F}$:
\begin{equation}
\myxymatrix{
  P\ar@{->}[d] & & & & \widehat{\Omega_1\sSU(2)}\ar@{->}[d]\\
 \sqcup_{a,b} U_a\cap U_b \ar@<-.5ex>[r] \ar@<.5ex>[r]& ~\sqcup_a U_a\ar@{->>}[d] & & & \Omega\sSU(2) \ar@<-.5ex>[r] \ar@<.5ex>[r] & P_0\sSU(2)\ar@{->>}[d]^{\dpar}\\
 & \sSU(2)& & & & \sSU(2)}
\end{equation}
Here the horizontal double arrows are the obvious maps from $U_a\cap U_b$ to $U_a$ or $U_b$ as well as the projection from a parameterized loop split at $\tau=\tfrac12\in[0,1]$ onto the two resulting based paths. Moreover, $P$ and $\widehat{\Omega_1\sSU(2)}$ are principal $\sU(1)$-bundles over $\sqcup_{a,b} U_a\cap U_b$ and $\Omega\sSU(2)$, respectively.

The string 2-group structure on $\CCG_{\rm F}$ in the skeletal model requires additional elements in the Segal--Mitchison group cohomology complementing the 3-cocycle describing $\CCG_{\rm F}$ geometrically. The loop space model, however, is specified canonically by pointwise multiplication and the canonical product on the Kac--Moody central extension $\widehat{\Omega_1\sSU(2)}$. Moreover, the loop space model yields a Lie 2-group which is unital and associative, albeit based on infinite-dimensional spaces. 

We will not need any further details on the Lie 2-group models and therefore we can directly move on to their Lie 2-algebras.

\subsection{The string Lie 2-algebra}\label{ssec:string_Lie_2_algebra}

A method of differentiating Lie 2-groups to Lie 2-algebras was given by {\v S}evera~\cite{Severa:2006aa}. This procedure yields corresponding Lie 2-algebras in the form of 2-term $L_\infty$-algebra, cf.~\cite{Baez:2003aa} and appendices~\ref{app:B} and~\ref{app:C}. In~\cite{Demessie:2016ieh}, {\v S}evera's method was used to differentiate the skeletal model, and the result is the {\em string Lie 2-algebra}
\begin{equation}
 \astring_{\rm sk}(3)\ =\ \big(~\FR~\xrightarrow{~0~} \asu(2)~\big)
\end{equation}
with non-trivial products
\begin{equation}
\begin{aligned}
 \mu_2&:\asu(2)\wedge \asu(2)\rightarrow \asu(2)~,~~~&\mu_2(x_1,x_2)&=[x_1,x_2]~,\\
 \mu_3&:\asu(2)\wedge \asu(2)\wedge \asu(2)\rightarrow \FR~,~~~&\mu_3(x_1,x_2,x_3)&=(x_1,[x_2,x_3])~,
\end{aligned}
\end{equation}
where $(-,-)$ is the appropriately normalized Killing form on $\asu(2)$. This Lie 2-algebra and closely related ones were first considered in~\cite{Baez:2003aa}. It was named the string Lie 2-algebra since it was shown to integrate to a 2-group model of the string group~\cite{Henriques:2006aa,Baez:2005sn}.

The fact that $\mu_1$ is trivial\footnote{This also informs the name \textit{skeletal}, as an $L_\infty$-algebra with trivial $\mu_1$ corresponds to a category where source and target maps agree, cf.~\cite{Baez:2003aa}.} might seem very restrictive at first glance. Note, however, that any Lie 2-algebra is categorically equivalent to one with $\mu_1=0$~\cite{Baez:2003aa}.

The loop space model, on the other hand, can also be differentiated in a straightforward fashion, since it corresponds to a strict Lie 2-group. The result is 
\begin{equation}
 \astring_{\hat \Omega}(3)\ =\ \big(~\Omega \asu(2)\oplus \FR\xrightarrow{~\mu_1~} P_0\asu(2)~)~,
\end{equation}
where $\mu_1$ is the concatenation of the projection $\widehat{\Omega_1\asu(2)}\cong \Omega\asu(2)\oplus \FR\rightarrow \Omega\asu(2)$ with the embedding $\Omega\asu(2)\embd P_0\asu(2)$ as closed based paths. The remaining non-trivial products are
\begin{equation}
 \begin{aligned}
  \mu_2&:P_0\asu(2)\wedge P_0\asu(2)\rightarrow P_0\asu(2)~,~~~\hspace{2cm}\mu_2(\gamma_1,\gamma_2)=[\gamma_1,\gamma_2]~,\\
  \mu_2&:P_0\asu(2)\otimes(\Omega\asu(2)\oplus\FR)\rightarrow \Omega\asu(2)\oplus\FR~,~~~\\
  &\hspace{4cm}\mu_2\big(\gamma,(\lambda,r)\big)=\left([\gamma,\lambda]\; ,\; -2\int_0^1 \dd\tau \left(\gamma(\tau),\dder{\tau}\lambda(\tau)\right)\right)~.
 \end{aligned}
\end{equation}

The categorical equivalence between both Lie 2-algebras was shown in~\cite{Baez:2005sn}. Explicitly, we have the morphisms\footnote{See appendix~\ref{app:D} for definitions and notation.} $\Phi=(\phi_0,\phi_1)$ and $\Psi=(\psi_0,\psi_1)$ with 
\begin{equation}
\astring_{\rm sk}(3)\xrightarrow{~\Phi~}\astring_{\hat\Omega}(3)\xrightarrow{~\Psi~}\astring_{\rm sk}(3)~.
\end{equation}
The chain maps $\phi_0$ and $\psi_0$ are given in the diagram
\begin{equation}\label{eq:skeletal_to_loop_morphism_1}
\myxymatrix{
 \FR \ar@{->}[d] \ar@{^{(}->}[r] & \Omega \asu(2)\oplus \FR \ar@{->}[d] \ar@{->}^-{{\rm pr}_\FR}[r] & \FR \ar@{->}[d] \\
 \asu(2) \ar@{^{(}->}^{\cdot f(\tau)}[r]& P_0\asu(2) \ar@{->}^{\dpar}[r] & \asu(2)
}
\end{equation}
where ${\rm pr}_\FR$ is the obvious projection, $\dpar:P_0\asu(2)\rightarrow \asu(2)$ is the endpoint evaluation and $\cdot f(\tau):\asu(2)\rightarrow P_0\asu(2)$ is the embedding of $x_0\in\asu(2)$ as the straight line $x(\tau)=x_0 f(\tau)$, where $f:[0,1]\rightarrow \FR$ is a smooth function with $f(0)=0$ and $f(1)=1$. The maps $\phi_1$ and $\psi_1$ read as 
\begin{equation}\label{eq:skeletal_to_loop_morphism_2}
 \phi_1(x_1,x_2)=\big([x_1,x_2](f(\tau)-f^2(\tau)),0\big)~,~~~
 \psi_1(x_1,x_2)=\int_0^{1}\dd \tau~ (\dot x_1,x_2)-(x_1,\dot x_2)~.
\end{equation}
Clearly, $(\Psi\circ \Phi)_0$ is the identity map, and using the composition of morphisms of Lie 2-algebras~\eqref{eq:comp_L2_morphisms}, we readily verify that $\Psi\circ \Phi=\id_{\astring_{\rm sk}}$. On the other hand, there is a 2-morphism $\chi:\Phi\circ \Psi\rightarrow \id_{\astring_{\hat \Omega}}$ encoded in a map
\begin{equation}\label{eq:def_chi}
 \chi:P_0\asu(2)\rightarrow \Omega\asu(2)\oplus \FR~,~~~\chi(\gamma)=(\gamma-f(\tau)\dpar\gamma,0)~,
\end{equation}
see appendix~\ref{app:D} for the relevant definitions. Further details are found in~\cite[Lemma 37]{Baez:2005sn}. Thus, $\astring_{\rm sk}(3)$ and $\astring_{\hat \Omega}(3)$ are equivalent as Lie 2-algebras. All this readily generalizes from $\asu(2)$ to arbitrary Lie algebras $\frg$ with a preferred element $\mu_3\in H^3(\frg,\FR)$.

\subsection{Further arguments for using the string 2-group}\label{ssec:arguments_for_string}

Our geometric picture in section~\ref{ssec:geometric_picture} strongly suggests that the correct gauge 2-group for an interesting non-abelian self-dual string solution should be a Lie 2-group model of the string group $\sString(3)$. There is a number of further reasons for this, which we shall now summarize.

First of all, in the string theory picture of monopoles as D1-branes ending on D3-branes, the Lie algebra $\asu(2)$ arises as part of the Lie algebra of endomorphisms acting on the Hilbert space of the fuzzy 2-sphere~\cite{Myers:1999ps}. The corresponding M-theory lift is expected to involve a fuzzy 3-sphere, cf.~\cite{Saemann:2012ex}. A fully satisfactory quantization of the 3-sphere as a multisymplectic manifold has not yet been given, but one would expect it to follow along the lines developed in~\cite{Bunk:2016rta}. There, it was shown that the string 2-group, and correspondingly its Lie 2-algebra, acts very naturally on the 2-Hilbert spaces arising in higher geometric quantization.

A very detailed discussion of further arguments from a string theory perspective is found in~\cite[Section~2.4]{Fiorenza:2012tb}. In particular, it has been shown in~\cite{Aschieri:2004yz} that the non-abelian principal 3-bundle arising from multiple coinciding M5-branes involves in general the group $\widehat{\Omega \sE_8}$ included in $\sString(\sE_8)$. Since non-abelian self-dual strings should be described by configurations of coinciding M5-branes, we are again led to a string 2-group model. Closely linked to this observation is the fact that the supergravity $C$-field can be regarded as a string structure~\cite{Fiorenza:2012tb}.

Finally, recall from the ADE-classification of (2,0)-theories that we expect a (2,0)-theory for any simply laced Lie algebras. For these, we can readily construct a corresponding string Lie 2-algebra and its integrating string Lie 2-group.

An obvious question is now how the string Lie 2-algebra of $\asu(2)$ is related to 3-Lie algebras, i.e.~the gauge structure underlying the BLG M2-brane model~\cite{Bagger:2007jr,Gustavsson:2007vu}. As shown in~\cite{Palmer:2012ya}, 3-Lie algebras are special cases of strict Lie 2-algebras. In particular, the 3-Lie algebra called $A_4$ is equivalent to the Lie 2-algebra 
\begin{equation}
 A_4~=~\big(~\au(2)\xrightarrow{~\mu_1~} \asu(2)\oplus\asu(2)~\big)~,
\end{equation}
where $\mu_1=0$ and $\mu_2$ is given by the commutator on $\asu(2)\oplus \asu(2)$ as well as the left and right actions of the $\asu(2)$ on $\au(2)$. It may be interesting to note that the string Lie 2-algebra $\astring_{\rm sk}(3)$ is categorically equivalent to the Lie 2-algebra
\begin{equation}
 \widetilde\astring_{\rm sk}(3)~=~\big(~\au(2)\cong\FR\oplus \asu(2)\xrightarrow{~(0,\id)~} \asu(2)\oplus\asu(2)~\big)~,
\end{equation}
see appendix~\ref{app:D} for details on equivalences of Lie 2-algebras. That is, both Lie 2-algebras can be based on the same graded vector space, but they differ in their products. Moreover, since both $A_4$ and $\astring_{\rm sk}(3)$ are skeletal (i.e.~$\mu_1=0$), it is quite obvious from the explicit form of Lie 2-algebra equivalences that they cannot be equivalent as Lie 2-algebras.

\section{Higher gauge theory and categorical equivalence}

We now come to the mathematical framework that we will use to derive the self-dual string equation.

\subsection{Higher gauge theory}\label{ssec:higher_gauge_theory}

A higher gauge theory is a dynamical principle on kinematical data living on a higher principal bundle. The full picture involves categorified spaces and groups, but we can restrict ourselves to local data. Therefore the higher analogue of Lie algebras will suffice in the following. For a definition of these {\em $L_\infty$-algebras}, we refer to appendix~\ref{app:B}. 

We thus start from an $n$-term $L_\infty$-algebra or Lie $n$-algebra\footnote{We use both terms interchangeably, cf.~appendix~\ref{app:B}.} $\sL=(\sL_0\xleftarrow{~\mu_1~}\sL_1 \xleftarrow{~\mu_1~}\dots\xleftarrow{~\mu_1~}\sL_{n-1})$, which one should think of as the higher analogue of the Lie algebra of the structure group in a principal bundle. In order to generalize connections from Lie algebra valued 1-forms to a collection of $L_\infty$-algebra valued $i$-forms, $i=1,\dots,n$ we can now use the following observation that will also become very useful later in our discussion. The tensor product of an $L_\infty$-algebra $\sL=\bigoplus_{k\in \RZ}\sL_k$ with the de~Rham complex on some manifold $M$ carries again an $L_\infty$-algebra structure, see e.g.~\cite{Jurco:2014mva}. The underlying graded vector space reads as
\begin{equation}
 \hat \sL=\Omega^\bullet(M)\otimes \sL=\bigoplus_{k\in\NN_0} \hat \sL_{-k}\ewith \hat \sL_{-k}=\bigoplus_{i-j=k} \Omega^i(M)\otimes \sL_j~,
\end{equation}
where the total degree is the $L_\infty$-degree minus the form degree. In this convention, the exterior derivative $\dd$ is of degree~$-1$ and $k$-forms with values in $\sL_0$ are of degree~$-k$.\footnote{Unfortunately, one always has to employ slightly awkward grading conventions when working simultaneously with $L_\infty$-algebras, differential forms and N$Q$-manifolds.} The higher products are defined as the linear span of the following products of elements $\omega_i\in\Omega^\bullet(M)$ and $\ell_i\in \sL$ of homogeneous degrees:
\begin{equation}\label{eq:tensor_mui}
 \hat \mu_i(\omega_1\otimes \ell_1,\dots,\omega_i\otimes \ell_i)\coloneqq \begin{cases} \dd \omega_1 \otimes \ell_1+(-1)^{|\omega_1|}\omega_1\otimes \mu_1(\ell_1)&\efor i=1~,\\
 \pm (\omega_1\wedge \dots \wedge \omega_i)\otimes \mu_i(\ell_1,\dots,\ell_i)&\qquad\mbox{else}~.
                                       \end{cases}
\end{equation}
The sign in the last equation is the Koszul sign convention\footnote{Interchanging two odd elements requires inserting a minus sign.} from permuting the  $\omega_i$ past the $\ell_i$ and the $\mu_i$.

A natural equation on a differential graded Lie algebra $(\frg,\dd)$ is the Maurer--Cartan equation $\dd a+\tfrac12[a,a]=0$ for $a\in \frg$ with $|a|=-1$. This generalizes in an obvious manner to the {\em homotopy Maurer--Cartan equation} for an element $\ell$ in some $L_\infty$-algebra $\sL$ with $|\ell|=-1$,
\begin{equation}\label{eq:hMC_equations_general}
 \sum_{k\in\NN}\frac{(-1)^{\frac{k(k+1)}{2}+1}}{k!}\mu_k(\ell,\dots,\ell)=0~.
\end{equation}
We will call an element $\ell$ satisfying this equation a {\em homotopy Maurer--Cartan element} of $\sL$.

A (reasonably small) gauge transformation between two homotopy Maurer--Cartan elements $\ell_1$ and $\ell_2$ in $\sL$ is given by a {\em homotopy} between them. That is, a homotopy Maurer--Cartan element $\hat \ell\in \Omega^\bullet([0,1])\otimes \sL$ such that $\hat\ell(0)=\ell_1$ and $\hat\ell(1)=\ell_2$. The homotopy Maurer--Cartan equations on $\hat \ell$ determine in particular $\der{t}\hat \ell|_{t=0}$. The latter can be identified with the change due to an infinitesimal gauge transformation, $\delta \ell$, while the gauge parameters $\lambda$ are given by the differential forms $\iota_{\der{t}} \hat\ell|_{t=0}$. Note that $\lambda$ is of degree~0, and we arrive at the formula
\begin{equation}\label{hMC_gauge_trafos}
 \delta \ell\ \coloneqq \ \sum_{k\in\NN} \frac{(-1)^{\frac{k(k+1)}{2}+1}}{(k-1)!}\mu_k(\ell,\dots,\ell,\lambda)~.
\end{equation}

The homotopy Maurer--Cartan equations~\eqref{eq:hMC_equations_general} describe a generalization of flat connections, but we shall be mostly interested in the non-flat case. We will call an element of $\Omega^\bullet\otimes \sL$ for some $L_\infty$-algebra $\sL$ an $\sL$-valued {\em higher connection} and the corresponding left-hand side of the homotopy Maurer--Cartan equation~\eqref{eq:hMC_equations_general} its {\em higher curvatures}. 

A Lie 2-algebra valued higher connection reads as $A+B\in \hat \sL_{-1}$ with $A\in \Omega^1(M)\otimes \sL_0$ and $B\in \Omega^2(M)\otimes \sL_{1}$. From the homotopy Maurer--Cartan equations~\eqref{eq:hMC_equations_general}, we can read off the corresponding higher curvatures
\begin{equation}\label{eq:naive_curvatures}
\begin{aligned}
 \CF&\coloneqq \dd A+\tfrac12 \mu_2(A,A)+\mu_1(B)~~~\in~ \Omega^2(M)\otimes \sL_0~,\\
 H&\coloneqq \dd B+\mu_2(A,B)+\tfrac{1}{3!}\mu_3(A,A,A)~~~\in~ \Omega^3(M)\otimes \sL_{1}~,
\end{aligned}
\end{equation}
which take values in $\hat \sL_{-2}$. From equation~\eqref{hMC_gauge_trafos}, we obtain the gauge transformations
\begin{equation}\label{eq:naive_gauge_trafos}
 \begin{aligned}
  \delta A&=\dd \alpha-\mu_1(\Lambda)+\mu_2(A,\alpha)~,\\
  \delta B&=\dd \Lambda+\mu_2(A,\Lambda)+\mu_2(B,\alpha)-\tfrac12\mu_3(A,A,\alpha)~,
 \end{aligned}
\end{equation}
which are parameterized by $\alpha\in\Omega^0(M)\otimes \sL_0$ and $\Lambda\in \Omega^1(M)\otimes \sL_{1}$. On curvatures, the gauge transformations act as
\begin{equation}\label{eq:naive_gauge_trafos_curvatures}
 \begin{aligned}
  \delta \CF &= \mu_2(\CF,\alpha)~,\\
  \delta H &=\mu_2(H,\alpha)+\mu_2(\CF,\Lambda)-\mu_3(\CF,A,\alpha)~.
 \end{aligned}
\end{equation}
Note that the $\mu_i$ here act only on the gauge structure of the fields and gauge parameters. They are {\em not} the $\hat \mu_i$ from~\eqref{eq:tensor_mui}.

\subsection{Higher gauge theory from differential graded algebras}\label{ssec:hgt_dgas}

Let us also give a different description of higher gauge theory which will be very useful for our subsequent discussion. This approach is a generalization of ideas by Atiyah~\cite{Atiyah:1957} partially due to~\cite{Bojowald:0406445,Kotov:2010wr,Gruetzmann:2014ica} and, to its full extent, due to~\cite{Sati:2008eg}. In this framework, connections and curvatures are described in terms of morphisms of differential graded algebras.

We shall need the definition of the Chevalley--Eilenberg algebra ${\rm CE}(\sL)$ of a Lie $n$-algebra $\sL$, as given in appendix~\ref{app:C}. This differential graded algebra sits inside the {\em Weil algebra} ${\rm W}(\sL)$ of $\sL$, which is the Chevalley--Eilenberg algebra of $T[1]\frg$ with differential $Q_{\rm W}=Q_{\rm CE}+\sigma$, where $\sigma$ is the nilquadratic grade-shift operation which anticommutes with $Q_{\rm CE}$,
\begin{equation}
 \sigma^2=0~,~~~\sigma \circ Q_{\rm CE}=-Q_{\rm CE}\circ \sigma~,
\end{equation}
and satisfies the graded Leibniz rule.

A {\em morphism of differential graded algebras} from ${\rm W}(\sL)$ to the de~Rham complex\linebreak $(\Omega^\bullet(M),\dd)={\rm W}(M)$ encodes now the local data for a connection and a curvature for structure Lie $n$-algebra $\sL$. Moreover, infinitesimal gauge transformations are encoded in homotopies between such morphisms for which the curvature vanishes along the homotopy direction.

As an illustrative example, consider an ordinary Lie algebra $\frg$. Its Chevalley--Eilenberg algebra ${\rm CE}(\frg)=\CC^\infty(\frg[1])$ is generated by coordinates $\xi^\alpha$ on $\frg[1]$ of degree~1 with differential $Q_{\rm CE}=-\tfrac12 f^\alpha_{\beta\gamma}\xi^\beta\xi^\gamma\der{\xi^\alpha}$. Its Weil algebra ${\rm W}(\frg)$ is generated by coordinates $\xi^\alpha$ and $\zeta^\alpha=\sigma \xi^\alpha$ on $T[1]\frg[1]$ of degree~1 and 2, respectively, with differential $Q_{\rm W}=Q_{\rm CE}+\sigma$ acting according to
\begin{equation}
 \begin{aligned}
   Q_{\rm W}\xi^\alpha&=-\tfrac12 f^\alpha_{\beta\gamma}\xi^\beta\xi^\gamma+\zeta^\alpha~,\\
   Q_{\rm W}\zeta^\alpha&=-\sigma Q\xi^\alpha=-f^\alpha_{\beta\gamma}\xi^\beta\zeta^\gamma~.
 \end{aligned}
\end{equation}
A morphism $\Theta$ from ${\rm W}(\frg)$ to $(\Omega^\bullet(M),\dd)=W(M)$ is fixed by the images of the coordinate functions, and we have
\begin{equation}
 \Theta:\xi^\alpha \mapsto A^\alpha \eand \Theta:\zeta^\alpha \mapsto F^\alpha~,
\end{equation}
where $A^\alpha$ and $F^\alpha$ are one- and two-forms, respectively. Compatibility with the differentials implies that
\begin{equation}
 \dd A^\alpha=-\tfrac12 f^\alpha_{\beta\gamma}A^\beta A^\gamma+F^\alpha\eor F=\dd A+\tfrac12 [A,A]
\end{equation}
as well as
\begin{equation}
 \dd F^\alpha=-f^\alpha_{\beta\gamma}A^\beta F^\gamma\eor \nabla F=\dd F+[A,F]=0~.
\end{equation}
Note that $F=(\dd\circ\Theta-\Theta\circ Q_{\rm CE})\xi^\alpha$ can also be seen as the failure of the map $\Theta$ to be a morphism of differential graded algebras on ${\rm CE}(\frg)$.

To obtain gauge transformations, we extend $M$ to $M\times I$, $I=[0,1]$ and impose flatness along $I$. Let $t$ be the coordinate along $I$. The condition $\iota_{\der{t}} F=0$ implies that
\begin{equation}
 \delta A\coloneqq \left.\der{t}\hat A\right|_{t=0}=\dd \alpha+[A,\alpha]~,
\end{equation}
where we split $\hat A=\alpha\dd t+A$.

The same discussion is readily extended to the case of Lie 2-algebras (as well as Lie $n$-algebras), and reproduces formulas~\eqref{eq:naive_curvatures} and~\eqref{eq:naive_gauge_trafos}.

Contained in the Weil algebra ${\rm W}(\sL)$ is the differential graded algebra of {\em invariant polynomials}, ${\rm inv}(\sL)$, whose images under $\Theta$ yield all the topological invariants. Explicitly, we have a complex
\begin{equation}
 {\rm inv}(\sL)~\embd~{\rm W}(\sL)~\rightarrow {\rm CE}(\sL)~,
\end{equation}
and elements $p$ of ${\rm inv}(\sL)$ are polynomials in the fiber coordinates of $T[1]\frg[1]$ satisfying $Q_{\rm W}p=0$. In the case of an ordinary Lie algebra as discussed above, an example of an invariant polynomial is $\kappa_{\alpha\beta}\zeta^\alpha \zeta^\beta$, where $\kappa_{\alpha\beta}$ is the Killing form on $\frg$. The topological invariant resulting from the image of $\Theta$ is simply the second Chern character $\tr(F^\dagger\wedge F)$.

\subsection{Categorical equivalence}\label{ssec:categorical_equivalence}

Clearly, we would expect field equations for higher connections to be transparent to categorical equivalences. In particular, we would think that for two equivalent Lie 2-algebra $\sL$ and $\tilde \sL$, there is an isomorphism between gauge equivalence classes of solutions to well-formed field equations for higher connections. Let us discuss this issue in some more detail.

Consider morphisms\footnote{See again appendix~\ref{app:D} for definitions and notation.} $\Phi=(\phi_0,\phi_1)$ and $\Psi=(\psi_0,\psi_1)$ underlying an equivalence of the Lie 2-algebras $\sL$ and $\tilde \sL$,
\begin{equation}
 \myxymatrix{ \sL \ar@/^1.5ex/[r]^{\Phi} & \tilde \sL \ar@/^1.5ex/[l]^{\Psi}}~.
\end{equation}
These induce morphisms between the corresponding Weil algebras ${\rm W}(\sL)$ and ${\rm W}(\tilde \sL)$, from which we derive morphisms $\Gamma$, $\Delta$ between local connective structures via the commutative diagram
\begin{equation}
 \xymatrixcolsep{4cm}
 \xymatrixrowsep{1.4cm}
 \myxymatrix{
    \Omega^\bullet(M) \ar@/_3ex/[d]_{\Gamma} & {\rm W}(\sL) \ar[l]_-{\Theta=(A+B,F+H)}\ar@/_3ex/[d]_{\Psi^*} \\
    \Omega^\bullet(M) \ar@/_3ex/[u]_{\Delta} & {\rm W}(\tilde \sL) \ar[l]_-{\tilde \Theta=(\tilde A+\tilde B,\tilde F+\tilde H)}\ar@/_3ex/[u]_{\Phi^*}
    }
\end{equation}

Let us give some details of this picture. Let $\tau_\alpha$ and $t_a$ form a basis on $\sL_0$ and $\sL_1$, respectively, and introduce corresponding coordinates $\xi^\alpha, b^a$, which we assign degrees~1 and 2, respectively, to match the N$Q$-picture. Let $(\tilde \tau_\mu,\tilde t_m)$ and $(\tilde \xi^\mu, \tilde b^m)$ be the corresponding basis vectors and coordinates on $\tilde \sL$. Then $\Phi$ reads as
\begin{equation}
 \phi_0(\tau_\alpha)=\phi_\alpha^\mu \tilde \tau_\mu~,~~~\phi_0(t_a)=\phi_a^m\tilde t_m~,~~~\phi_1(\tau_\alpha,\tau_\beta)=\phi^m_{\alpha\beta}\tilde t_m
\end{equation}
for some coefficients $\phi_\alpha^\mu$, $\phi_a^m$ and $\phi^m_{\alpha\beta}$. At the level of the Chevalley--Eilenberg algebra, we have
\begin{equation}
 \tilde \xi^\mu\mapsto \phi_\alpha^\mu \xi^\alpha\eand \tilde b^m\mapsto \phi_a^mb^a+\tfrac12 \phi^m_{\alpha\beta} \xi^\alpha\xi^\beta~.
\end{equation}
To lift this to a morphism between the Weil algebras of $\tilde \sL$ and $\sL$, we introduce additional coordinates $(\zeta^\alpha,c^a)$ and $(\tilde \zeta^\mu, \tilde c^m)$ of degrees~(2,3). The lift of $\Phi$ then necessarily reads as
\begin{equation}
\begin{aligned}
 \tilde \xi^\mu&\mapsto \phi_\alpha^\mu \xi^\alpha~,~~~&\tilde b^m&\mapsto \phi_a^mb^a+\tfrac12 \phi^m_{\alpha\beta} \xi^\alpha\xi^\beta~,\\
 \tilde \zeta^\mu &\mapsto \phi_\alpha^\mu \zeta^\alpha~,~~~&\tilde c^m&\mapsto \phi_a^m c^a+\phi^m_{\alpha\beta}\zeta^\alpha \xi^\beta~.
\end{aligned}
\end{equation}

From this map, we can read off the action of the morphisms on the gauge potentials and their curvatures,
\begin{equation}\label{eq:image_Phi_connection}
\begin{aligned}
 \tilde A+\tilde B&\coloneqq \Phi(A+B)=\phi_0(A)+\phi_0(B)+\tfrac12\phi_1(A,A)~,\\
 \tilde \CF+\tilde H&\coloneqq \Phi(\CF+H)=\phi_0(\CF)+\phi_0(\CH)+\phi_1(\CF,A)~.
\end{aligned}
\end{equation}
Note that this map behaves well under composition of Lie 2-algebra morphisms. Also, the gauge parameters $\alpha+\Lambda$ as introduced above are mapped to $\Phi(\alpha+\Lambda)$ with
\begin{equation}
 \tilde \alpha+\tilde \Lambda=\Phi(\alpha+\Lambda)=\phi_0(\alpha+\Lambda)+\phi_1(\alpha+\Lambda,A+B)=\phi_0(\alpha)+\phi_0(\Lambda)+\phi_1(\alpha,A)~.
\end{equation}
We then have
\begin{equation}
 \delta_{\tilde \alpha+\tilde \Lambda}(\tilde A+\tilde B)=\widetilde{\delta_{\alpha+\Lambda}(A+B)}+\phi_1(\alpha,\CF)~.
\end{equation}

Let us now consider an important example of a higher gauge theory, the self-duality equation for the 3-form curvature of a higher connection on $M=\FR^{1,5}$, $H=*H$. At least for $\sL=(\au(1)\rightarrow *)$, this equation is linked to the description of M5-branes. Again, under a morphism $\Phi$ of $L_\infty$-algebras, the underlying higher connection $A+B$ will be mapped to $\tilde A+\tilde B$ as in~\eqref{eq:image_Phi_connection}, and for $H$, we have
\begin{equation}
 \tilde H=\phi_0(H)+\phi_1(\CF,A)~.
\end{equation}
That is, for $H=*H$ to be transparent under equivalence of Lie 2-algebra, we have to complement it with the {\em fake flatness} condition $\CF=0$. This condition is also responsible for rendering a higher parallel transport of strings invariant under surface reparametrizations~\cite{Baez:2004in}.

The same result evidently holds also for the dimensional reduction of the self-duality equation to the non-abelian self-dual string equation in four dimensions, see e.g.~\cite{Saemann:2011nb,Saemann:2012uq}. Fake flatness, however, is too strong a condition since it allows for gauge transformations to the abelian case~\cite{Demessie:2016ieh}, which we can readily see as follows. The obvious lift of the abelian self-dual string equation $H=*\dd \varphi$ to higher gauge theory with skeletal string Lie 2-algebra $\astring_{\rm sk}(3)$ takes the same form 
\begin{equation}\label{eq:sdseom1}
 H\coloneqq \dd B+\tfrac{1}{3!}\mu_3(A,A,A)=* \dd \varphi~.
\end{equation}
Fake flatness $\CF=F=\dd A+\tfrac12 \mu_2(A,A)=0$ implies that $A$ is equivalent to the trivial connection by a finite gauge transformation, cf.~\cite{Demessie:2016ieh}. The self-dual string equation~\eqref{eq:sdseom1} then reduces to the abelian one.

The same statement is true in the loop space picture. Assume that $\CF=\dd A+\tfrac12 \mu_2(A,A)+\mu_1(B)=0$. This is a gauge invariant statement and for convenience, we gauge transform by $\Lambda=A-f(\tau)\dpar A$ such that $A=f(\tau)\dpar A\eqqcolon f(\tau)A_0$ and $f(\tau)>0$ for all $\tau\in[0,1]$. Then
\begin{equation}
 F=\dd A+\tfrac12 \mu_2(A,A)=f(\tau)\dd A_0+\tfrac12 (f(\tau))^2[A_0,A_0]~.
\end{equation}
Since $\CF=0$, $F$ is in the image of $\mu_1$, which implies that $\dd A_0+\tfrac12 [A_0,A_0]=0$ and $A_0$ is pure gauge. It follows that also $A$ is of the form $A=g(\tau)\dd (g(\tau))^{-1}$ and thus pure gauge. Again, $\CF=0$ cannot be part of the equations of motion of a truly non-abelian self-dual string.

Essentially this problem was encountered before, in a general discussion of non-abelian higher gauge theories based on strict Lie 2-groups~\cite{Ho:2012nt} as well as in a first approach to using higher gauge theory to describe non-abelian self-dual strings~\cite{Palmer:2013haa}. The new loophole to this problem is a reformulation of the above expressions of higher gauge theory in a way which is more suitable for the description of string and M-theory, as we shall explain now.

\subsection{Reformulation: Twisted higher gauge theory}\label{ssec:twist}

In the following, we briefly outline the arguments of~\cite{Sati:2009ic,Fiorenza:2010mh,Fiorenza:2012tb}, which explain how to cast our above formulas for a non-abelian higher gauge theory on the worldvolume of M5-branes in a more natural way. The upshot of these arguments will be presented in section~\ref{ssec:sds_eom}, and it is safe to jump directly to this subsection.

Our above formulas were not suited for accommodating appropriate anomaly cancellation condition arising in M-theory. At first sight, one might assume that this condition is irrelevant; after all, anomaly cancellation is a global issue and we are interested in higher gauge theory over the contractible space $\FR^4$. Note, however, that our solutions correspond to topological configurations, and it is not surprising that these turn out to imply a one-point compactification of $\FR^4$ to $S^4$. 

First, recall that the first Chern class of a $\sU(n)$-bundle is the obstruction for reducing the structure group of the underlying principal bundle to $\sSU(n)$. In terms of stacks $\sB\sG$ of principal $\sG$-bundles, this is reflected in the following homotopy pullback:
\begin{equation}
\myxymatrix{
 \sB\sSU(n) \ar@{->}[r] \ar@{->}[d]& {*} \ar@{->}[d]\\
 \sB\sU(n) \ar@{->}[r]^{c_1} & \sB\sU(1)}
\end{equation}
Correspondingly, we can describe principal $\sSU(n)$-bundles alternatively as principal $\sU(n)$-bundles $P$ with a choice of trivialization of the determinant line bundle $c_1(P)$. This generalizes to principal bundles with connections.

In the case of supergravity on an M5-brane boundary~\cite{Horava:1996ma,Witten:1996hc}, the kinematical data is contained in a spin connection and an $\sE_8$ gauge field. Anomaly cancellation requires that the first fractional Pontryagin class $\tfrac12 p_1$ of the first equals twice the canonical 4-class of the latter. The two characteristic classes are maps
\begin{equation}
 \sB\sSpin(n)\xrightarrow{~\tfrac12 p_1~}\sB^3\sU(1)\eand \sB\sE_8\xrightarrow{~a~}\sB^3\sU(1)~,
\end{equation}
and fit into the homotopy pullback
\begin{equation}\label{eq:hom_pullback}
\myxymatrix{
 \sB\sString^a(n) \ar@{->}[r] \ar@{->}[d]& \sB\sE_8 \ar@{->}[d]^a\\
 \sB\sSpin(n) \ar@{->}[r]^{\tfrac12 p_1} & \sB^3\sU(1)}
\end{equation}
where $\sB^3\sU(1)$ is the (higher) stack of 2-gerbes, circle 3-bundles or principal $\sB^2\sU(1)$-bundles. The anomaly cancellation condition modifies the Bianchi identity $\dd H=\tfrac{1}{2}\mu_3(\dd A,A,A)$ coming from~\eqref{eq:naive_curvatures} to 
\begin{equation}
 \dd H=(F_A,F_A)-2(F_{\tilde A},F_{\tilde A})~,
\end{equation}
where $A$ is the spin connection and $\tilde A$ is the $\sE_8$ gauge field.

For our discussion, it is sufficient to ignore the $\sE_8$ gauge field and put $a=0$. Following the above picture, we now describe a principal $\sString(n)$-bundle as a principal $\sSpin(n)$-bundle with a trivialization of a principal $\sB^2\sU(1)$-bundle (with connection)~\cite{Fiorenza:2012tb}. A precise derivation of this picture is found in~\cite[Section~6.3]{Fiorenza:2010mh} or~\cite[Section~7.1.6.3]{Schreiber:2013pra}. To develop the appropriate notions of gauge potentials and connections as discussed in section~\ref{ssec:hgt_dgas}, we switch to the differential graded algebras of the involved Lie 3-algebras. We arrive at the commutative diagram
\begin{equation}\label{eq:comm_diag_modified_Weil}
 \xymatrixcolsep{11pc}
 \myxymatrix{ 
 {\rm CE}(\aspin(n)) & {\rm CE}(\sB^2\au(1)) \ar@{->}[l]_{\mu\coloneqq -\tfrac{1}{3!}\mu_{\alpha\beta\gamma}\xi^\alpha\xi^\beta\xi^\gamma}\\
 {\rm W}(\aspin(n)) \ar@{->}[u] & {\rm W}(\sB^2\au(1)) \ar@{->}[u] \ar@{->}[l]_{{\rm cs}\coloneqq \kappa_{\alpha\beta}\xi^\alpha \zeta^\beta-\tfrac{1}{3!}\mu_{\alpha\beta\gamma} \xi^\alpha\xi^\beta\xi^\gamma}\\
 {\rm inv}(\aspin(n)) \ar@{->}[u] & {\rm inv}(\sB^2\au(1)) \ar@{->}[u]\ar@{->}[l]_{\tfrac12 p_1\coloneqq \kappa_{\alpha\beta}\zeta^\alpha \zeta^\beta}
 }
\end{equation}
where $\xi^\alpha$ and $\zeta^\alpha$ are coordinates on ${\rm W}(\aspin(n))$, introduced in section~\ref{ssec:categorical_equivalence}.

It is now convenient to replace $\aspin(n)$ with the equivalent Lie 3-algebra $\widehat{\astring}_{\rm sk}(n)$,
\begin{equation}
 \widehat{\astring}_{\rm sk}(n)\ =\ \sB\au(1)\rightarrow \astring_{\rm sk}(n)\ =\ \au(1) \xrightarrow{~\id~} \au(1) \xrightarrow{~0~} \frg~,
\end{equation}
with nontrivial brackets
\begin{equation}
 \mu_2(x_1,x_2)=[x_1,x_2]~,~~~\mu_3(x_1,x_2,x_3)=(x_1,[x_2,x_3])~,~~~\mu_1(s)=s
\end{equation}
for $x_i\in \frg$ and $s\in \au(1)$. The equivalence of $\aspin(n)$ with $\widehat{\astring}_{\rm sk}(n)$ follows from the $L_\infty$-algebra quasi-isomorphism
\begin{equation}
 \myxymatrix{ 
  {*} \ar@{->}[r] \ar@{^{(}->}[d]^{\phi_0} & {*} \ar@{^{(}->}[d]^{\phi_0}\ar@{->}[r] & \frg \ar@{->}[d]^{\phi_0=\id}\\
  \FR \ar@{->}[r]^{\id} & \FR \ar@{->}[r]^0 & \frg
 }~~~~\ewith\phi_2(x_1,x_2,x_3)=(x_1,[x_2,x_3])~,
\end{equation}
cf.~appendix~\ref{app:D}. Following the observation about the first Chern class above, we can now describe principal $\sString(n)$-bundles as principal 3-bundles with structure Lie 3-algebra $\widehat{\astring}_{\rm sk}(n)$, together with additional data trivializing the image of $\tfrac12 p_1$. Note that a related description has appeared in~\cite{Waldorf:2009uf}.

One subtle point in this picture, however, is that we need to modify the structure of the Weil algebra in~\eqref{eq:comm_diag_modified_Weil}, employing $\tilde{\rm W}(\widehat{\astring}_{\rm sk}(n))$, to guarantee that the diagram
\begin{equation}
 \xymatrixcolsep{3pc}
 \myxymatrix{ 
 {\rm CE}(\aspin(n)) & {\rm CE}(\widehat\astring_{\rm sk}(n)) \ar@{->}[l]_{\cong} & {\rm CE}(\sB^2\au(1)) \ar@{->}[l]\\
 {\rm W}(\aspin(n)) \ar@{->}[u] & {\rm \tilde W}(\widehat\astring_{\rm sk}(n)) \ar@{->}[u] \ar@{->}[l]_{\cong} & {\rm W}(\sB^2\au(1)) \ar@{->}[u] \ar@{->}[l]\\
 {\rm inv}(\aspin(n)) \ar@{->}[u] & {\rm inv}(\widehat\astring_{\rm sk}(n)) \ar@{->}[u] \ar@{->}[l]_{=}& {\rm inv}(\sB^2\au(1)) \ar@{->}[u]\ar@{->}[l]
 }
\end{equation}
commutes. In coordinates $\xi^\alpha,\zeta^\alpha,b,c,k,l$ of degrees~$1,2,2,3,3,4$, respectively, the differential $Q$ on $\tilde{\rm W}(\widehat{\astring}_{\rm sk}(n))$ acts as follows:
\begin{equation}
\begin{aligned}
 Q\xi^\alpha&=-\tfrac12 f^\alpha_{\beta\gamma}\xi^\beta\xi^\gamma+\zeta^\alpha~,~~~&Q\zeta^\alpha&=-f^\alpha_{\beta\gamma}\xi^\beta\zeta^\gamma~,\\
 Qb&={\rm cs}+c-k~,~~~&Qc&=l-\tfrac12 p_1~,\\
 Qk&=l~,~~~&Ql&=0~,
\end{aligned}
\end{equation}
cf.~\cite{Sati:2009ic}. The expressions ${\rm cs}$ and $\tfrac12 p_1$ are defined in diagram~\eqref{eq:comm_diag_modified_Weil}. Applying the formalism of section~\ref{ssec:hgt_dgas} yields the following local description over a contractible manifold $M$~\cite{Sati:2009ic}. We have gauge potentials
\begin{equation}
 A\in \Omega^1(M)\otimes \frg~,~~~B\in \Omega^2(M)\otimes \au(1)~,~~~C\in \Omega^3(M)\otimes \au(1)
\end{equation}
with curvatures
\begin{equation}
 F=\dd A+\tfrac12[A,A]~,~~~H=\dd B+C-(A,\dd A)-\tfrac{1}{3}(A,[A,A])~,~~~G=\dd C~,
\end{equation}
satisfying the Bianchi identities
\begin{equation}
 \nabla F=0~,~~~\dd H=G-(F,F)~,~~~\dd G=0~.
\end{equation}
Since ${\rm CE}(\widehat\astring_{\rm sk}(n))$ does not change, $H$ is no longer the failure of $\Theta$ to be a morphism of differential graded algebras at the level of ${\rm CE}(\widehat\astring_{\rm sk}(n))$. In the following, we shall refer to the above potentials and curvatures with $C=0$ as higher gauge theory with structure $L_\infty$-algebra $\astring_{\rm sk}^T(n)$.

We note that similar curvatures arose from a non-abelian version of the tensor hierarchies in supergravity~\cite{Samtleben:2011fj}, see also~\cite{Palmer:2013pka} for further relations to higher gauge theory.

Gauge transformations are readily derived via homotopies which are partially flat, as detailed in section~\ref{ssec:hgt_dgas}, and we obtain
\begin{equation}
\begin{aligned}
\delta A &= \dd\alpha + \mu_2(A,\alpha)~,\\
\delta B &= \dd \Lambda - \Sigma +(\alpha,F) -\tfrac12 \mu_3(A,A,\alpha)~,\\
\delta C &= \dd \Sigma~,
\end{aligned}
\end{equation}
which are now parameterized by $\alpha\in\Omega^0(\FR^4)\otimes \asu(2)$, $\Lambda \in\Omega^1(\FR^4)\otimes \FR$ and $\Sigma\in\Omega^2(\FR^4)\otimes\FR$. Under these transformations the curvatures transform as
\begin{equation}
\label{eq:gaugetrafo_FHG}
\delta F = \mu_2(F,\alpha)~,~~~\delta H = 0~,~~~\delta G = 0~.
\end{equation}
We note that the gauge invariance of $H$ allows us to write down a gauge invariant self-duality equation in six dimensions. Studying this equation is beyond the scope of this paper, but certainly of great interest.

When working with the loop space model $\astring_{\hat \Omega}(3)$ we will have to implement an analogous twist, which we shall explain in section~\ref{ssec:sds_eom2}

\section{Non-abelian self-dual strings}
In this section, we derive the self-dual string equations and present explicit solutions for both the skeletal as well as the loop model.

\subsection{The abelian self-dual string}

Consider a single flat M5-brane trivially embedded into flat Minkowski space $\FR^{1,10}$ such that the time direction is contained in the M5-brane's worldvolume. The presence of the M5-brane breaks some of the isometries (and their superpartners) of $\FR^{1,10}$ as well as some of the gauge symmetries of the 3-form background field in supergravity. The corresponding collective modes yield the $(2,0)$ tensor multiplet in six dimensions~\cite{Adawi:1998ta}, which consists of 5~scalars, a 2-form $B$ with self-dual curvature $H=\dd B=* H$ together with sixteen fermionic partners. We shall focus on the 2-form field $B$.

An M2-brane may end on an M5-brane with a one-dimensional boundary~\cite{Strominger:1995ac,Townsend:1996em}, which is the M-theory analogue of the fact that a string and a D2-brane can end on a D4-brane. If the remaining direction of the M2-brane is perpendicular to the worldvolume of our flat M5-brane from above, we arrive at the {\em self-dual string soliton}~\cite{Howe:1997ue}. The latter is governed by the equation
\begin{equation}\label{eq:ab_SDS}
 H=\dd B=* \dd \varphi
\end{equation}
on the part of the worldvolume $\FR^{4}$ of the M5-brane which it does not share with the M2-brane. In this $\FR^4$, the boundary of the M2-brane is a point $x_0\in \FR^4$. Note that this equation is a dimensional reduction of the self-duality equation $H=*H$ from $\FR^{1,5}$ to $\FR^4$ with the scalar field $\varphi$ identified with the components of $B$ along the reduced directions. Also, a further dimensional reduction to $\FR^3$ yields the abelian Bogomolny monopole equations $F=*\dd \varphi$.

From the Bianchi identity $\dd H=0$, we learn that $\varphi$ is a harmonic functions on $\FR^4$. Therefore, interesting solutions will be singular at a point $x_0$. For a single self-dual string at $x_0$, the solution is
\begin{equation}\label{eq:sol_ab_SDS}
 \varphi=\frac{1}{(x-x_0)^2}~,
\end{equation}
and the concrete expression for the $B$-field, which is singular along Dirac strings going from the origin through opposite poles of $S^3$ to infinity, can be found (up to its radial dependence) e.g.~in~\cite{Nepomechie:1984wu}. Because equation~\eqref{eq:ab_SDS} is linear in both $B$ and $\varphi$, we can form linear combination of solutions to obtain new solutions. That is, the abelian self-dual strings do not interact.

This is fully analogous to the Dirac monopole. In the case of the latter, we can obtain non-singular and interacting configurations by considering non-abelian generalizations. Our aim is the construction of corresponding non-singular and interacting self-dual strings.

\subsection{The self-dual string equations in the skeletal case}\label{ssec:sds_eom}

After our discussion in section~\ref{ssec:twist}, it is now straightforward to write down the 3-form part of the non-abelian self-dual string equation on $\FR^4$ for the two models of the string Lie 2-algebra. In the case of $\astring_{\rm sk}^T(3)$, we have kinematical data consisting of fields 
\begin{equation}\label{eq:sds_fc_sk}
 A\in \Omega^1(\FR^4)\otimes \asu(2)~,~~~B\in \Omega^2(\FR^4)\otimes \au(1)~,~~~\varphi\in \Omega^0(\FR^4)\otimes \au(1)~,
\end{equation}
satisfying
\begin{equation}\label{eq:sds_eom_sk_1}
 H\coloneqq \dd B-(A,\dd A)-\tfrac{1}{3}(A,[A,A])=*\dd \varphi~.
\end{equation}
Next, note that the Bianchi identity leads to 
\begin{equation}\label{eq:Bianchi}
 *\dd H=-*(F,F)=\square \varphi~,
\end{equation}
and therefore the Higgs field $\varphi$ is determined by the second Chern character, which captures instantons on $\FR^4$. Since knowing the Higgs field should suffice to describe the self-dual string modulo gauge invariance, it is natural to replace fake flatness $F=0$ with the instanton equation
\begin{equation}\label{eq:sds_eom_sk_2}
 F=*F~.
\end{equation}
This result is also in agreement with a different point of view. In the six-dimensional $\CN=(1,0)$ supersymmetric model of~\cite{Samtleben:2011fj}, the BPS equation leads to $\square \varphi= -*(F,*F)$~\cite{Akyol:2012cq}. This BPS equation follows from our equation~\eqref{eq:Bianchi}, if it is supplemented with the instanton equation $F=*F$. We shall discuss the implications of our choice of gauge structure for the (1,0)-model in section~\ref{sec:10model}.

As a first consistency check, note that by putting $A=0$, our equations~\eqref{eq:sds_eom_sk_1} and~\eqref{eq:sds_eom_sk_2} reduce to the abelian self-dual string equation~\eqref{eq:ab_SDS}.

Another consistency check that we can immediately perform is the reduction from M2-branes ending on M5-branes to D2-branes ending on D4-branes. That is, we dimensionally reduce $\FR^4$ along an M-theory direction, say $x^4$. The resulting kinematical data consists of the following fields
\begin{equation}
\begin{aligned}
 \breve{A}_1&\in \Omega^1(\FR^3)\otimes \asu(2)~,~~~&\breve{\varphi}_1&\in \Omega^0(\FR^3)\otimes \au(1)~,~~~&\breve{B}&\in \Omega^2(\FR^3)\otimes \au(1)~,\\
 \breve{A}_2&\in \Omega^1(\FR^3)\otimes \au(1)~,~~~&\breve{\varphi}_2&\in\Omega^0(\FR^3)\otimes \asu(2)~.
\end{aligned}
\end{equation}
Our equations~\eqref{eq:sds_eom_sk_1} and~\eqref{eq:sds_eom_sk_2} reduce to the following expressions:
\begin{equation}
\begin{aligned}
*\nabla \breve \varphi_2 &= \dd \breve A_1 + \tfrac12\mu_2(\breve A_1,\breve A_1)\eqqcolon  \breve F~,\\
0&=\dd \breve B  - (\breve A_1,\dd \breve A_1) - \tfrac{1}{3}(\breve A_1,[\breve A_1,\breve A_1])\eqqcolon  \breve H_1~,\\
*\dd \breve \varphi_1 &= \dd \breve A_2 - (\breve A_1,\dd \breve\varphi_2)-(\breve\varphi_2,\dd\breve A_1)-(\breve A_1, [\breve A_1, \breve\varphi_2])\eqqcolon  \breve H_2~.
\end{aligned}
\end{equation}
Here the first equation is just the monopole equation on $\FR^3$ for connection $\breve A_1$ and Higgs field $\breve \varphi_2$. The second equation can be satisfied by choosing an appropriate $\breve B$. This is possible by Poincar\'e's lemma, since the form to be canceled by $\dd \breve B$ is a top form on $\FR^3$, hence closed. The third equation can be rewritten as
\begin{equation}
*\dd \breve \varphi_1- \dd \breve A_2=-2(\breve \varphi_2,\breve F)+\dd (\breve \varphi_2,\breve A_1)=-*\dd(\breve \varphi_2,\breve \varphi_2)+\dd (\breve \varphi_2,\breve A_1)~,
\end{equation}
where we used $\breve F=*\nabla \breve \varphi_2 $. This is clearly solved by
\begin{equation}
 \breve A_2=-(\breve \varphi_2,\breve A_1)\eand \breve \varphi_1=-(\breve \varphi_2,\breve \varphi_2)~.
\end{equation}
Altogether, the dimensional reduction of our self-dual string equations~\eqref{eq:sds_eom_sk_1} and~\eqref{eq:sds_eom_sk_2} leads to the Bogomolny monopole equations on $\FR^3$, as expected from string theory.

\subsection{The self-dual string equations for the loop space model}\label{ssec:sds_eom2}

Next, let us consider the corresponding equations for the model $\astring_{\hat \Omega}(3)$. Here, the kinematical data is given by fields 
\begin{equation}\label{eq:sds_fc_loop}
 A\in \Omega^1(\FR^4)\otimes P_0\asu(2)~,~~~B\in \Omega^2(\FR^4)\otimes (\Omega\asu(2)\oplus \FR)~,~~~\varphi\in \Omega^0(\FR^4)\otimes (\Omega\asu(2)\oplus \FR)~.
\end{equation}
Analogously to the skeletal case, we also need to modify the original 3-form curvature $H=\dd B+\mu_2(A,B)$ to render $H$ gauge invariant. The correct twist to $\astring^T_{\hat \Omega}(3)$ is now given by
\begin{equation}\label{eq:twisted_H_loop}
\begin{aligned}
 \CF &\coloneqq \dd A+\tfrac12\mu_2(A,A)+\mu_1(B)~,\\
 H&\coloneqq \dd B+\mu_2(A,B)-\kappa(A,\CF)~,
\end{aligned}
\end{equation}
where 
\begin{equation}
\begin{aligned}
 \kappa:P_0\asu(2)\times P_0\asu(2)&\rightarrow \Omega\asu(2)\oplus \FR~,\\
 \kappa(\gamma_1,\gamma_2)&\coloneqq \left(\chi([\gamma_1,\gamma_2])\;,\;2\int_0^1\dd \tau (\dot \gamma_1,\gamma_2)\right)
\end{aligned}
\end{equation}
and $\chi(\gamma)=(\gamma-f(\tau)\dpar \gamma,0)$ was defined in equation~\eqref{eq:def_chi}. Note that for $\gamma_1$ or $\gamma_2$ a loop, $\kappa(\gamma_1,\gamma_2)=\mu_2(\gamma_1,\gamma_2)$. This modifies the gauge transformations as follows:
\begin{equation}
 \begin{aligned}
  \delta A&=\dd \alpha+\mu_2(A,\alpha)-\mu_1(\Lambda)~,\\
  \delta B&=\dd \Lambda+\mu_2(A,\Lambda)-\mu_2(\alpha,B)+\kappa(\alpha,\CF)~.
 \end{aligned}
\end{equation}
The 2- and 3-form curvatures~\eqref{eq:twisted_H_loop} then transform according to
\begin{equation}
 \begin{aligned}
  \delta \CF&=\mu_2(\CF,\alpha)+\mu_1(\kappa(\alpha,\CF))~,\\
  \delta H&=0~.
 \end{aligned}
\end{equation}

Having a gauge invariant $H$ at our disposal, we readily write down a suitable set of equations of motion:
\begin{equation}\label{eq:sds_eq_loop}
\CF =*\CF~,~~~ H=*\nabla \varphi~,~~~ \mu_1(\varphi)=0~.
\end{equation}
Note that the third equation\footnote{This condition is trivially satisfied in the skeletal case.} is necessary and sufficient to render the second one gauge invariant.\footnote{As remarked earlier, one can also write down a gauge invariant self-duality equation in six dimensions.} The appropriate Bianchi identities read as
\begin{equation}
\nabla\CF = \mu_1(\kappa(A,\CF))+\mu_1(H)~,~~~ \nabla H =\mu_2(A,H)-\kappa(\CF,\CF)~.
\end{equation}

The dimensional reduction to monopoles is now accomplished by restricting to the endpoint in path space, $\dpar \CF$, and projecting onto $\FR$ in $\Omega\asu(2)\oplus \FR$, where we recover the skeletal situation.

In fact, it is not hard to see that equations~\eqref{eq:sds_eq_loop} are categorically equivalent to~\eqref{eq:sds_eom_sk_1} and~\eqref{eq:sds_eom_sk_2} in the sense that gauge equivalence classes of solutions to each of these equations are in one-to-one correspondence with gauge equivalence classes of solutions of the respective other ones. This correspondence is established by the map~\eqref{eq:image_Phi_connection} of higher gauge potentials induced by the morphisms of Lie 2-algebras underlying the equivalence
\begin{equation}
 \myxymatrix{ \astring_{\rm sk}(3) \ar@/^3ex/[r]^{\Phi} & \astring_{\hat \Omega}(3)\ar@/^3ex/[l]^{\Psi}}
\end{equation}
as defined in~\eqref{eq:skeletal_to_loop_morphism_1} and~\eqref{eq:skeletal_to_loop_morphism_2}. More explicitly, given a solution $(A_{\rm sk},B_{\rm sk},\varphi_{\rm sk})$ to~\eqref{eq:sds_eom_sk_1} and~\eqref{eq:sds_eom_sk_2}, one readily verifies that 
\begin{equation}\label{eq:loop_space_from_sk}
 A_{\hat \Omega}=A_{\rm sk}f(\tau)~,~~~B_{\hat \Omega}=(\tfrac12[A_{\rm sk},A_{\rm sk}](f(\tau)-f^2(\tau)),B_{\rm sk})\eand \varphi_{\hat \Omega}=(0,\varphi_{\rm sk})
\end{equation}
is a solution to~\eqref{eq:sds_eq_loop}. Here, $f:[0,1]\rightarrow \FR$ is some smooth function with $f(0)=0$ and $f(1)=1$. Gauge transformations of $(A_{\rm sk},B_{\rm sk},\varphi_{\rm sk})$ parameterized by $(\alpha_{\rm sk},\Lambda_{\rm sk})$ are mapped to gauge transformations of $(A_{\hat \Omega},B_{\hat \Omega},\varphi_{\hat \Omega})$ parameterized by
\begin{equation}
 \alpha_{\hat \Omega}=\alpha_{\rm sk}f(\tau)\eand \Lambda_{\hat \Omega}=\big([\alpha_{\rm sk},A_{\rm sk}](f(\tau)-f^2(\tau))\;,\;\Lambda_{\rm sk}\big)~.
\end{equation}

Conversely given a solution $(A_{\hat \Omega},B_{\hat \Omega},\varphi_{\hat \Omega})$, it is straightforward to check that 
\begin{equation}
 A_{\rm sk}=\dpar A_{\hat \Omega}~,~~~B_{\rm sk}=\pr_{\FR} B_{\hat \Omega}+\int^1_0\dd \tau~(A_{\hat \Omega},\dot{A}_{\hat \Omega})~,~~~\varphi_{\rm sk}=\pr_{\FR}\varphi_{\hat \Omega}
\end{equation}
is a solution to~\eqref{eq:sds_eom_sk_1} and~\eqref{eq:sds_eom_sk_2}. Moreover, gauge transformations parameterized by $(\alpha_{\hat \Omega},\Lambda_{\hat \Omega})$ are mapped to gauge transformations parameterized by 
\begin{equation}
 \alpha_{\rm sk}=\dpar \alpha_{\hat \Omega}\eand \Lambda_{\rm sk}=\pr_{\FR} \Lambda_{\hat \Omega}+\int_0^1\dd\tau~(\dot \alpha_{\hat\Omega},A_{\hat\Omega})-(\alpha_{\hat\Omega},\dot A_{\hat\Omega})~.
\end{equation}

\subsection{Bogomolny bound}

Recall that both the instanton and monopole equations can be derived as equations for the Bogomolny bound of a suitable action principle. The same is true for our non-abelian self-dual string equations. This will also lead to an identification of the appropriate topological charge.

For simplicity, we restrict ourselves to the skeletal case $\astring_{\rm sk}^T(n)$. We then have the following obvious action functional of higher Yang--Mills--Higgs theory:
\begin{equation}
 S=\int_{\FR^4} H\wedge *H+\dd \varphi\wedge *\dd \varphi+(F,*F)~,
\end{equation}
where $F$, $H$ and $\varphi$ are the 2- and 3-form curvature as well as the Higgs field introduced in the above sections.
For $\varphi=0$, this action was given before in~\cite{Baez:2002jn} in a more general context, where it was, however, not gauge invariant. Since $\delta H=0$ for the skeletal string Lie 2-algebra (as well as for the twisted strict Lie 2-algebra), this problem does not arise here. We can recast this action in the following form:
\begin{equation}
 S=\int_{\FR^4} (H-*\dd\varphi)\wedge *(H-*\dd\varphi)-2H\wedge \dd \varphi+\tfrac12\big((F-*F),*(F-*F)\big)+\big(F,F\big)~.
\end{equation}
As expected, the minimum of this action is given by solutions to our self-dual string equations
\begin{equation}
 H=*\dd\varphi~,~~~F=*F~,
\end{equation}
and for such solutions, the action is given by the topological invariants
\begin{equation}
 S=-2 \int_{\FR^4} H\wedge \dd\varphi+\int_{\FR^4} \big(F,F\big)=-2 \int_{\FR^4} H\wedge \dd\varphi+\int_{S^3_\infty} H~,
\end{equation}
where we used the Bianchi identity $\dd H=(F,F)$. 

\subsection{The elementary solution}\label{ssec:solution}

Let us now come to the explicit form of the elementary solution, starting with the case of the skeletal algebra $\astring_{\rm sk}^T(3)$. The relevant field content is~\eqref{eq:sds_fc_sk} and we wish to solve
\begin{equation}\label{eq:eom_sds_2}
 H\coloneqq\dd B+(A,\dd A)+\tfrac{1}{3}(A,[A,A])=*\dd \varphi\eand F\coloneqq \dd A+\tfrac12 [A,A]=*F~.
\end{equation}
We start from the elementary instanton solution and a trivial 2-form potential,
\begin{equation}\label{eq:sol_sk_1}
 A_\mu(x)=-\di\frac{\,\eta_{\mu\nu}^i\,\sigma_i\, (x^\nu-x_0^\nu)}{\rho^2+(x-x_0)^2}~,~~~B(x)=0~,
\end{equation}
where $\sigma_i$ are the Pauli matrices satisfying $[\sigma_i,\sigma_j]=2\di \eps_{ijk}\sigma_k$ and $\eta^i_{\nu\kappa}$ are the 't Hooft symbols, which form a basis of self-dual 2-forms on $\FR^4$. The variables $x_0\in \FR^4$ and $\rho\in \FR$ denote the position and the size of the elementary instanton. The inner product $(-,-)$ on $\asu(2)$ is the one appropriately normalized Killing form,
\begin{equation}
 (x,y)={\rm tr}(x^\dagger y)\ewith (\di \sigma_i,\di \sigma_j)=(\sigma_i,\sigma_j)=\delta_{ij}~.
\end{equation}
With these conventions, we find that
\begin{equation}\label{eq:sol_sk_2}
 \varphi(x)=\frac{(x-x_0)^2+2\rho^2}{\big((x-x_0)^2+\rho^2\big)^2}
\end{equation}
completes the solution.

Let us now perform the obvious consistency checks on our solution~\eqref{eq:sol_sk_1} and~\eqref{eq:sol_sk_2}. First of all, it is evident that this solution is non-singular on all of $\FR^4$, which sets it apart from the abelian solution~\eqref{eq:sol_ab_SDS}. In the limit $|x|\rightarrow \infty$, however, $\varphi\sim \frac{1}{x^2}$, which is the solution to the abelian self-dual string.

The moduli of our elementary solution are the same as those of the instanton: the position $x_0$, the size parameter $\rho$ as well as a global gauge transformation $g\in \sSU(2)$. The size parameter is the Goldstone mode arising from the break down of  conformal invariance of the instanton equation $F=*F$ by choosing a specific solution~\eqref{eq:sol_sk_1}. 

For the loop space model $\astring^T_{\hat\Omega}(3)$, we can simply use categorical equivalence to translate our solution~\eqref{eq:image_Phi_connection}. Here, the relevant field content is listed in~\eqref{eq:sds_fc_loop} with the corresponding curvatures~\eqref{eq:twisted_H_loop} and equations of motion~\eqref{eq:sds_eq_loop}.

Using~\eqref{eq:loop_space_from_sk} on the skeletal solution~\eqref{eq:sol_sk_1}, we obtain the potentials
\begin{equation}
\begin{aligned}
A_\mu(x) &= -\di \frac{\,\eta_{\mu\nu}^i\,\sigma_i\, (x^\nu-x_0^\nu)}{\rho^2+(x-x_0)^2}f(\tau)~,\\
B_{\mu\nu}(x) &= \left( -2\,\di \,\epsilon_{ijk}\,\sigma_k\frac{\,\eta_{\mu\kappa}^i\,(x-x_0)^\kappa\, \eta_{\nu\lambda}^j\,(x-x_0)^\lambda}{(\rho^2+(x-x_0)^2)^2}(f(\tau)-f^2(\tau))\; ,\;0\right)~
\end{aligned}~,
\end{equation}
where $f:[0,1]\rightarrow \FR$ is again a smooth function with $f(0)=0$ and $f(1)=1$, as well as the Higgs field
\begin{equation}
\varphi(x) = \left(0,\frac{(x-x_0)^2+2\rho^2}{\big((x-x_0)^2+\rho^2\big)^2}\right)~.
\end{equation}
These indeed form a solution to equations~\eqref{eq:sds_eq_loop}, as expected. Conversely, we recover the skeletal case from the inverse morphism of gauge potentials.

\section{The global picture: string structures}\label{sec:global_picture}

While our discussion so far is in principle consistent on flat $\FR^4$, the full geometric picture has a remaining issue. For simplicity, we shall discuss this problem for the skeletal string Lie 2-algebra; the corresponding discussion in the strict case follows rather trivially.

The fact that we consider instantons on $\FR^4$ suggests that we are working on a compactification $M$ of $\FR^4$ such as e.g.~$S^4$. In this case, the first fractional Pontryagin class $\tfrac12 p_1=(F,F)$ is not trivial in $H^4(M,\RZ)$. This, however, would be a requirement for our gauge potentials to live on a principal 2-bundle corresponding to a string structure as defined in~\cite{Killingback:1986rd,Stolz:2004aa,Redden:2006aa,Waldorf:2009uf,Sati:2009ic}, cf.\ also the discussion in section~\ref{ssec:twist}.

There are two rather obvious loopholes to this problem. First, we can extend the structure $L_\infty$-algebra $\astring^T_{\rm sk}(3)$ in such a way that the additional degrees of freedom compensate the instanton contribution to the first Pontryagin class. Second, we can turn on the $\sE_8$ gauge degrees of freedom available in the moduli stack of supergravity $C$-field connections~\cite{Fiorenza:2012tb}, i.e.~let $a\neq 0$ in~\eqref{eq:hom_pullback}, to achieve the same.

The first solution is rather natural, as the following argument shows. Recall that D1-branes ending on D3-branes form a fuzzy funnel, in which points of the worldvolume of the D1-branes polarize into fuzzy 2-spheres~\cite{Myers:1999ps}. The double cover of the isometry group of the 2-sphere then has an action on the Hilbert space arising from geometric quantization. The same group is then the gauge group of the worldvolume theory on the D1-branes.

In the case of M2-branes ending on M5-branes, one expects a polarization into fuzzy 3-spheres, which need to be quantized by a categorified version of geometric quantization, cf.~\cite{Bunk:2016rta}. We then expect the string group of the double cover of the isometries of $S^3$, namely $\sSpin(4)\cong \sSU(2)\times \sSU(2)$, to act on the categorified Hilbert space and to underlie the gauge structure on the M2-brane side. That is, we may want to replace $\astring_{\rm sk}(3)$ by $\astring_{\rm sk}(4)=\big(\FR\rightarrow \asu(2)\times \asu(2)\big)$, which also brings our equations closer to the M2-brane models of~\cite{Bagger:2007jr,Gustavsson:2007vu,Aharony:2008ug}. 

The Pontryagin classes with respect to both $\asu(2)$-factors add, and we can compensate the instanton $F_L$ in the left factor $\asu(2)_L$ with an anti-instanton $F_R$ in the right factor $\asu(2)_R$ to obtain $[\tfrac12 p_1]=0$ in $H^4(M,\RZ)$. Altogether, we arrive at the equations
\begin{equation}\label{eq:string_structure}
 F_L=*F_L~,~~F_R=-*F_R~,~~[\tfrac12 p_1]=[(F,F)]=[(F_L,F_L)+(F_R,F_R)]=0~,~~H=*\dd \varphi~.
\end{equation}
We note that an alternative way of arriving at equivalent data is to flip the sign of the Killing form on $\asu_R(2)$, leading to an indefinite metric on $\asu_L(2)\oplus \asu_R(2)$, which is precisely the gauge algebra underlying the simplest M2-brane model. In this case, both $F_L$ and $F_R$ are chosen self-dual.

From our previous results in section~\ref{ssec:solution}, we readily glean the following extended solution:
\begin{gather}
  A_{\mu,L}(x)=-\di\frac{\,\eta_{\mu\nu}^i\,\sigma_i\, (x^\nu-x_{0,L}^\nu)}{\rho^2_L+(x-x_{0,L})^2}~,~~~A_{\mu,R}(x)=-\di\frac{\,\bar \eta_{\mu\nu}^i\,\sigma_i\, (x^\nu-x_{0,R}^\nu)}{\rho^2_R+(x-x_{0,R})^2}~,\\
  B(x)=0~,~~~\varphi=\frac{(x-x_{0,L})^2+2\rho^2_L}{\big((x-x_{0,L})^2+\rho^2_L\big)^2}-\frac{(x-x_{0,R})^2+2\rho^2_R}{\big((x-x_{0,R})^2+\rho^2_R\big)^2}~,
\end{gather}
where the 't Hooft tensors $\etab^i_{\mu\nu}$ form a basis for anti-self-dual 2-forms in four dimensions. Note that the instanton and the anti-instanton do not have to have the same size $\rho$ nor do they have to be centered at the same point $x_0$. If all the moduli agree, then evidently $\varphi=0$ and thus $H=0$. 

The above data on $\FR^4=S^4\backslash\{\infty\}$ provides us now with a truly non-trivial and well-defined string structure on $S^4$. While we cannot present all the details of the complete picture (as e.g.~in terms of non-abelian differential cocycles; this is work in progress), we can see the non-triviality in the gauge connections. We have a principal $\sSpin(4)$-bundle with connection defined by $A=A_L+A_R$ over $S^4$ and the projections onto the two underlying $\sSU(2)$-bundles are topologically clearly non-trivial: carrying an instanton and an anti-instanton, their individual Pontryagin classes do not vanish and gauge transformations do not mix them. The other characteristic class relevant here is the 3-form curvature $H$. This should be understood as the sum of a Dixmier--Douady class $\dd B$ of an abelian gerbe (which, by itself, is necessarily trivial on $S^4$) and a coboundary ${\rm cs}(A)$ that trivializes a 2-gerbe with 4-form curvature $(F,F)$. Note that $H$ is gauge invariant and therefore isomorphisms of principal 2-bundles (which are gauge transformations) will not affect its value. It is thus indeed an invariant of string structures. As far as we are aware, this is the first explicit example of a non-trivial and truly non-abelian gerbe relevant to string or M-theory.

\section{The 6d superconformal field theory}\label{sec:10model}

Having clarified the gauge structure as well as the equations of motion of the non-abelian self-dual string, it is natural to ask about implications for a classical 6d superconformal field theory. In particular, one would like our equations to arise as the BPS equations of such a theory.

For convenience, we shall restrict ourselves in the following to the Lie 3-algebra\linebreak $\widehat{\astring}_{\rm sk}(n)$ that we constructed in section~\ref{ssec:twist} and which underlies our self-dual string equations; a categorically equivalent treatment of the loop space models should exist.

To formulate an action, we require the appropriate notion of an inner product, which is given by a cyclic structure on our gauge $L_\infty$-algebra. The Lie 3-algebra $\widehat{\astring}_{\rm sk}(n)$, however, does not carry such a structure. Therefore, we first need to minimally extend $\widehat{\astring}_{\rm sk}(n)$ to a Lie 3-algebra $\widehat{\astring}_\omega(n)$ with cyclic structure. It turns out that this minimal extension encodes in fact a $(1,0)$ gauge structure in the sense of~\cite{Palmer:2013pka}. That is, it is a suitable gauge structure for the $\CN=(1,0)$ superconformal field theory in six dimensions with action that was proposed in~\cite{Samtleben:2011fj}. This result is an extension of the observation made in~\cite{Palmer:2013pka} that the string Lie 2-algebra is a $(1,0)$ gauge structure.

\subsection{Cyclic Lie 3-algebra structure}\label{ssec:ext_Lie_3}

As briefly explained in appendix~\ref{app:C}, a cyclic structure on a $k$-term $L_\infty$-algebra or Lie $k$-algebra $\frg$ is most readily constructed from a symplectic form on the corresponding N$Q$-manifold $\frg[1]$. Note that given the N$Q$-manifold $\frg[1]$ of any Lie $k$-algebra $\frg$ (in particular with $\frg[1]$ not necessarily symplectic), we can double it to $T^*[k+1]\frg[1]$, which is concentrated in the same degrees and clearly symplectic. A vector field $Q$ on $T^*[k+1]\frg[1]$ is then found by extending that on $\frg[1]$ minimally, and the result corresponds to a doubled and cyclic Lie $k$-algebra $T^*[k-1]\frg$.

The N$Q$-manifold corresponding to $\widehat{\astring}_{\rm sk}(n)$ reads as
\begin{equation}
 \widehat{\astring}_{\rm sk}(n)[1]\ = \ \frg[1] \leftarrow \FR_r[2] \leftarrow \FR_p[3]~,
\end{equation}
where\footnote{Our construction works for arbitrary metric Lie algebra $\frg$.} $\frg=\aspin(3)$ and the subscripts will help us distinguish different copies of the real line and indicate the coordinates we will be using: $x^a$, $r$ and $p$ on the subspaces $\frg[1]$, $\FR_r[2]$ and $\FR_p[3]$, which are of degree~1, 2 and 3, respectively. In these coordinates, the vector field $Q$ reads as
\begin{equation}\label{eq:Q_hat_string}
 Q=-\frac12 f^a_{bc}x^bx^c\der{x^a}-\frac{1}{3!}f_{abc}x^ax^bx^c\der{r}+p\der{r}~.
\end{equation}
Doubling the N$Q$-manifold as sketched above, we have 
\begin{equation}
\widehat{\astring}_\omega(n)[1]\coloneqq \ \frg[1]\oplus\FR_q[1]\leftarrow \FR_r[2]\oplus\FR_s[2]\leftarrow \FR_p[3]\oplus \frg^*[3]~,
\end{equation}
where we coordinatize the subspaces $\frg[1]$, $\FR_q[1]$, $\FR_r[2]$, $\FR_s[2]$, $\FR_p[3]$, $\frg^*[3]$ by $x^a$, $q$, $r$, $s$, $p$, $y_a$, which have degrees~1, 1, 2, 2, 3, 3, respectively. The natural symplectic form of (N$Q$-)degree~4 reads as
\begin{equation}
 \omega=\dd x^a\wedge \dd y_a+\dd q\wedge \dd p+\dd r\wedge \dd s~,
\end{equation}
which induces the following Poisson bracket on $\CC^\infty(\widehat{\astring}_\omega(n)[1])$:
\begin{equation}
 \{f,g\}\coloneqq -f\overleftarrow{\der{y_a}}\overrightarrow{\der{x^a}} g- f\overleftarrow{\der{x^a}}\overrightarrow{\der{y_a}} g-f\overleftarrow{\der{p}}\overrightarrow{\der{q}} g- f\overleftarrow{\der{q}}\overrightarrow{\der{p}} g+ f\overleftarrow{\der{s}}\overrightarrow{\der{r}} g- f\overleftarrow{\der{r}}\overrightarrow{\der{s}} g~.
\end{equation}
The Poisson bracket allows us to work with the Hamiltonian function $\CQ$ of the vector field $Q=\{\CQ,-\}$, which is more convenient. The minimal Hamiltonian function which induces an extension of~\eqref{eq:Q_hat_string} reads as 
\begin{equation}
 \CQ=-\tfrac12 f^a_{bc}x^bx^cy_a-\tfrac{1}{3!}f_{abc}x^ax^bx^cs+sp~,
\end{equation}
where $f^a_{bc}$ are the structure constants of $\frg$. The corresponding Hamiltonian vector field is 
\begin{equation}
 Q=-\frac{1}{2} f^a_{bc}x^bx^c\der{x^a}-f^b_{ac}x^cy_b\der{y_a}+\frac{1}{2} f_{abc}x^bx^cs\der{y_a}-\frac{1}{3!}f_{abc}x^ax^bx^c\der{r}+p\der{r}+s\der{q}~.
\end{equation}
One readily checks $\{\CQ,\CQ\}=Q\CQ=0$, which is equivalent to $Q^2=0$.

Let us now translate back to the Lie 3-algebra picture. We shall use the same letters as above to denote elements of the various subspaces, but the grading is as indicated here:
\begin{equation}
 \widehat{\astring}_\omega(n)= \frg\oplus \FR_q\leftarrow \FR_r[1]\oplus \FR_s[1]\leftarrow \FR_p[2]\oplus \frg^*[2]~.
\end{equation}
The cyclic inner product on $\widehat{\astring}_\omega(n)$ now reads as 
\begin{multline}
 \langle x_1+q_1+r_1+s_1+p_1+y_1,x_2+q_2+r_2+s_2+p_2+y_2\rangle=\\y_1(x_2)+y_2(x_1)+p_1q_2+q_1p_2+r_1s_2+s_1r_2~,
\end{multline}
and the non-trivial higher products of the Lie 3-algebra are 
\begin{equation}
\begin{aligned}
 &\mu_1:\FR_s[1]\rightarrow \FR_q:~\mu_1(s)\coloneqq s~,\\
 &\mu_1:\FR_p[2]\rightarrow \FR_r[1]:~\mu_1(p)\coloneqq p~,\\
 &\mu_2:\frg\wedge \frg\rightarrow \frg:~ \mu_2(x_1,x_2)\coloneqq [x_1,x_2]~,\\
 &\mu_2:\frg\wedge \frg^*[2]\rightarrow \frg^*[2]:~ \mu_2(x,y)\coloneqq y([-,x])~,\\
 &\mu_3:\frg\wedge \frg\wedge \frg\rightarrow \FR_r[1]:~\mu_3(x_1,x_2,x_3)\coloneqq (x_1,[x_2,x_3])~,\\
 &\mu_3:\frg\wedge \frg\wedge \FR_s[1]\rightarrow \frg^*[2]:~\mu_3(x_1,x_2,s)\coloneqq (-,[x_1,x_2])s~.
\end{aligned}
\end{equation}

Next, we observe that the data available to us can be used to refine the Lie 3-algebra $\widehat{\astring}_\omega(n)$ to a $(1,0)$ gauge structure which we also denote by $\widehat{\astring}_\omega(n)$ and which can be used for the (1,0)-model of~\cite{Samtleben:2011fj}. For simplicity, we shall use the conventions and notation of~\cite{Palmer:2013pka} and abbreviate $\widehat{\astring}_\omega(n)=\fra_0\oplus \fra_{1}\oplus \fra_2$. Explicitly, we have maps 
\begin{equation}
 \begin{aligned}
  \sfh:~&\fra_{1}\rightarrow \fra_0~,~~~&\sfh(r+s)&=\mu_1(r+s)=s~,\\
  \sfg:~&\fra_{2}\rightarrow \fra_{1}~,~~~&\sfg(y+p)&=\mu_1(y+p)=p~,\\
  \sff:~&\fra_0\wedge \fra_0\rightarrow \fra_0~,~~~&\sff(x_1+q_1,x_2+q_2)&=\mu_2(x_1,x_2)=[x_1,x_2]~,\\
  \sfd:~&\fra_0\odot \fra_0\rightarrow \fra_{1}~,~~~&\sfd(x_1+q_1,x_2+q_2)&=\tfrac12(x_1,x_2)\in\FR_r[1]~,\\
  \sfb:~&\fra_{1}\otimes\fra_0\rightarrow \fra_{2}~,~~~&\sfb(r+s,x+q)&=(-,x)s\in\frg^*[2]~.
 \end{aligned}
\end{equation}

It is now easy to verify that these maps satisfy the necessary relations for a $(1,0)$ gauge structure admitting an action, which read as
\begin{equation*}
\begin{aligned}
 \sfh(\sfg(\lambda))&=0~,\\
 \sff(\sfh(\chi),\gamma)-\sfh(\sfd(\sfh(\chi),\gamma))&=0~,\\
 \sff(\gamma_{[1},\sff(\gamma_2,\gamma_{3]}))-\tfrac{1}{3}\sfh(\sfd(\sff(\gamma_{[1},\gamma_2),\gamma_{3]}))&=0~,\\
 \sfg(\sfb(\chi_1,\sfh(\chi_2)))-2\sfd(\sfh(\chi_1),\sfh(\chi_2))&=0~,\\ 
 \sum_\alpha\big(\big\langle\lambda,\sfh(\sfd(\gamma,\tau_\alpha))-\sff(\gamma,\tau_\alpha)\big\rangle\big)\langle\sfg(\tau^\alpha_*),\chi\rangle-\langle\sfg(\sfb(\sfg(\lambda),\gamma))),\chi\rangle&=0~,\\
 \langle \chi,\sfg(\lambda)\rangle-\langle \sfh(\chi),\lambda\rangle &=0~,\\
 \langle \chi,\sfd(\gamma_1,\gamma_2)\rangle-\tfrac12 \langle \sfb(\chi,\gamma_1),\gamma_2\rangle&=0~,\\
\end{aligned}
\end{equation*}
\begin{equation*}
\begin{aligned}
  \big\langle\sfd(\gamma_1,\gamma_{(2}),\sfd(\gamma_3,\gamma_{4)})\big\rangle&=0~,\\
2(\sfd(\sfh(\sfd(\gamma_1,\gamma_{(2})),\gamma_{3)})-\sfd(\sfh(\sfd(\gamma_2,\gamma_3)),\gamma_1))-2\sfd(\sff(\gamma_1,\gamma_{(2}),\gamma_{3)})+\sfg(\sfb(\sfd(\gamma_2,\gamma_3),\gamma_1))&=0~,\\
\big\langle \sfb(\chi,\sfh(\sfd(\gamma_1,\gamma_2)))+2\sfb(\sfd(\gamma_1,\sfh(\chi)),\gamma_3)-\sfb(\chi,\sff(\gamma_1,\gamma_3))-\sfb(\sfg(\sfb(\chi,\gamma_1)),\gamma_3),\gamma_2\big\rangle&\\
+\langle\sfb(\chi,\gamma_3),\sfh(\sfd(\gamma_1,\gamma_2))-\sff(\gamma_1,\gamma_2)\rangle&=0~,\\
\end{aligned}
\end{equation*}
where $\gamma_i,\gamma\in \fra_0$, $\chi\in \fra_{1}$ and $\lambda\in \fra_{2}$ and $\tau_\alpha$ is some basis of $\frg$ and $\tau^\alpha_*$ is the dual basis of $\frg^*[2]$.

\subsection{Action and BPS equations}

We can now proceed and specialize the action of~\cite{Samtleben:2011fj} to the $(1,0)$ gauge structure $\widehat{\astring}_\omega(n)$. We shall label all fields as in~\cite{Samtleben:2011fj}, only suppressing the gauge indices. The relevant field content is arranged in an $\CN=(1,0)$ vector multiplet and an $\CN=(1,0)$ tensor multiplet in six dimensions. The $\CN=(1,0)$ vector multiplet $(A,\lambda^i,Y^{ij})$ consists of a 1-form potential $A$, spinors $\lambda^i$ and (auxiliary) scalar field $Y^{ij}=Y^{ji}$ taking values in $\fra_0=\frg\oplus \FR$, where $i=1,2$ are indices for the vector representation of the R-symmetry group $\sSp(1)$. These indices can be raised and lowered by the $\sSp(1)$-invariant matrices $\eps^{ij}$ and $\eps_{ij}$. The $\CN=(1,0)$ tensor multiplet $(\varphi,\chi^i,B)$ consists of a 2-form potential $B$, a spinor $\chi$ and a scalar field $\varphi$ taking values in $\fra_{1}=\FR_r[1]\oplus \FR_s[1]$ and an (auxiliary) 3-form potential $C$, taking values in $\fra_{2}=\frg^*[2]\oplus \FR[2]$. The corresponding field strengths are given by
\begin{equation}
 \CF=\dd A+\tfrac12 [A,A]+\mu_1(B)\eand \CH=\dd B-(A,\dd A)-\tfrac13(A,[A,A])+\mu_1(C)~,
\end{equation}
together with the Bianchi identities
\begin{equation}
 \nabla \CF=\mu_1(\CH)\eand \nabla \CH=-(\CF,\CF)+\mu_1(\CG)
\end{equation}
for some 4-form field strength $\CG=\dd C+\dots$, where higher terms are suppressed, cf.~\cite{Samtleben:2011fj}. The action then reads as
\begin{equation}\label{eq:6d_action}
\begin{aligned}
 S=\int_{\FR^{1,5}}&\Big(-\tfrac{1}{8}\langle\dd \varphi, *\dd \varphi\rangle-*\tfrac12 \langle \bar \chi_i ,\dpars \chi^i\rangle+\tfrac{1}{16}\varphi_s\big((\CF,*\CF)-*4(Y_{ij},Y^{ij})+*8(\bar \lambda_i,\nablas\lambda^i)\big)\\
 &~~~-\tfrac{1}{96}\langle\CH,*\CH\rangle-\tfrac{1}{48}\CH_s\wedge*\langle \lambdab_i,\gamma_{(3)}\lambda^i\rangle-\tfrac14 \langle \lambdab_i,\CF\rangle\wedge *\gamma_{(2)}\chi^i_{s}+*\langle Y_{ij},\bar\lambda^i\rangle \chi^j_s\\
 &~~~-\mu_1(C)\wedge\CH_s+B_s\wedge(\CF,\CF)\Big)~.
\end{aligned}
\end{equation}
Here, $\gamma_{(n)}\coloneqq \frac{1}{n!}\gamma_{\mu_1\dots\mu_n} \dd x^{\mu_1}\wedge \dots \wedge \dd x^{\mu_n}$ and subscripts $s$ denote the components of the fields taking values in $\FR_s[1]$. The last line in~\eqref{eq:6d_action} is clearly a purely topological term, see~\cite{Samtleben:2011fj} for details. Self-duality of $\CH$ is imposed by hand in the form of the equation
\begin{equation}
 \CH-*\CH=-(\lambdab^i,\gamma_{(3)} \lambda_i)~.
\end{equation}

The general action of~\cite{Samtleben:2011fj} came with a few undesirable features. First, it was not clear which (1,0) gauge structures should be chosen. Second, the action generically contains a cubic interaction term of the scalar fields $\varphi$, leading to potential instabilities. Third, if the action is not free, it necessarily comes with an indefinite kinetic term for these scalar fields. While our choice of $(1,0)$ gauge structure solves the first two issues, the problem of the indefinite kinetic term is still present. It can possibly be addressed analogously to similar terms in M2-brane models based on Lorentzian 3-Lie algebras.

Considering the supersymmetry transformations, it is easy to extract BPS equations as done in~\cite{Akyol:2012cq,Akyol:2013ana}. Their results were given for a Lie algebra $\frg$ together with a representation $\CR$. Choosing $\CR$ to be the trivial representation $\FR$, the BPS equations contain in particular
\begin{equation}
 \square \varphi=*(\CF,*\CF)~,
\end{equation}
which is satisfied if our equations~\eqref{eq:sds_eom_sk_1} and~\eqref{eq:sds_eom_sk_2} are fulfilled. The self-duality equation here follows from the Killing spinor equation in the specific setting of~\cite{Akyol:2012cq,Akyol:2013ana}. We thus note that imposing self-duality in~\eqref{eq:sds_eom_sk_2} was the correct requirement for a BPS state in the $(1,0)$-model of~\cite{Samtleben:2011fj}.

\subsection{Reduction to four-dimensional super Yang--Mills theory}

The reduction of our self-dual string equations~\eqref{eq:sds_eom_sk_1} and~\eqref{eq:sds_eom_sk_2} to three dimensions was rather straightforward, which motivates an attempt at a corresponding reduction of the (1,0)-model to lower dimensions. While a reduction to five dimensions is not immediately obvious, a further reduction to $\CN=2$ super Yang--Mills theory in four dimensions arises directly: the action contains the term
\begin{equation}
 S=\int_{\FR^{1,5}}\tfrac{1}{16}\varphi_s\big((\CF,*\CF)-*4(Y_{ij},Y^{ij})+*8(\bar \lambda_i,\nablas\lambda^i)\big)~, 
\end{equation}
and performing a dimensional reduction along two space-like directions such that $\varphi_s$ gets a large expectation value will clearly yield a dominating $\CN=2$ super Yang--Mills part in the remaining four-dimensional Lagrangian. 

To justify such a large expectation value, recall the corresponding reduction for M2-branes. To reduce the BLG M2-brane model to super Yang--Mills theory, a dimensional reduction of one spatial direction transverse to the M2-branes is necessary. In~\cite{Mukhi:2008ux}, the authors conjecture that this amounts to the corresponding scalar field developing a vacuum expectation value proportional to the radius of the circle. A rough analogy in our case suggests that the expectation value developed by $\varphi_s$ should equal the area of the torus, which would satisfy the requirement. This, however, is very preliminary and we leave a full analysis in the context of M5-branes to future work.

A better string theory inspired picture may come from the NS5-D4 brane configurations studied in~\cite{Brunner:1997gk,Hanany:1997gh}. Slight generalisations of our theory seem to be reasonably good candidates for their description. In this picture, the scalar field $\varphi_s$ corresponds essentially to the distance of two parallel NS5-branes with D4-branes suspended between them. Thus, a large expectation value of $\varphi_s$ amounts to moving the two NS5-branes far apart, which should indeed lead to a decoupling of the tensor multiplet on the NS5-branes from the vector multiplet on the D4-branes and therefore produce four-dimensional super Yang--Mills theory. Clearly, again, much more work is needed to match our theory to string theory expectations, but what we have seen so far is encouraging.

\section*{Acknowledgements}

We would like to thank Branislav Jur{\v c}o, Michael Murray, Urs Schreiber, Konrad Waldorf and Martin Wolf for useful discussions. C.S.~was supported in part by the STFC Consolidated Grant ST/L000334/1 {\it Particle Theory at the Higgs Centre}. L.S.~was supported by an STFC PhD studentship.

\appendices

Below, we collect some definitions, results and hopefully helpful references. While appendix~\ref{app:A} is a rough overview over the categorification of the mathematical notions involved in the definition of principal bundles, appendices~\ref{app:B},~\ref{app:C} and~\ref{app:D} explain in detail various aspects of categorified gauge algebras. Potentially useful and more detailed introductions into some of the following material are found in~\cite{Baez:2010ya,Saemann:2016sis}.

\subsection{Categorified Lie groups, Lie algebras and principal bundles}\label{app:A}

Let $\CatMfd$ be the category of smooth spaces with smooth maps between them. A {\em 2-space} (e.g.~\cite{Baez:2004in}) is formally a category internal to\footnote{Note that $\CatMfd$ does not contain all pullbacks which leads to some technicalities which we can ignore.} $\CatMfd$. That is, a 2-space consists of two manifolds $M_1, M_0$ with smooth source and target maps $\sfs,\sft:M_1\rightrightarrows M_0$, a smooth embedding $\id:M_0\embd M_1$ and a smooth composition of morphisms $\circ: M_1\times_{M_0} M_1\rightarrow M_1$ satisfying the usual axioms of the structure maps of a category. {\em Morphisms between 2-spaces} can be defined as functors internal to $\CatMfd$, which are smooth maps between the morphism- and object-manifolds of the 2-spaces.\footnote{To be precise, one would immediately specialize from 2-spaces to Lie groupoids and regard these as differentiable stacks or objects in a 2-category with the much more general bibundles as morphisms, see e.g.~the discussion in~\cite{Demessie:2016ieh}. This point can safely be ignored.} Note that any manifold $M$ gives trivially rise to a 2-space $M\rightrightarrows M$ in which the only morphisms are the identity morphisms. Also, {\em Lie groupoids} are 2-spaces in which each morphism is invertible. 

In its simplest, strict form, a {\em Lie 2-group} is a category internal to $\CatGrp$, the category of groups. Such strict Lie 2-groups are most conveniently described as crossed modules of Lie groups~\cite{Baez:0307200}. More generally, a Lie 2-group is a Lie groupoid which is simultaneously a monoidal category and in which morphisms are strictly invertible and objects are weakly invertible with respect to the monoidal product~\cite{Baez:0307200}.

There is a higher analogue of Lie differentiation, which takes a {\em Lie 2-group} to a {\em Lie 2-algebra}~\cite{Baez:2003aa}. Strict Lie 2-groups differentiate to strict Lie 2-algebras, which are conveniently described by crossed modules of Lie algebras. More generally, a Lie 2-group differentiates to a 2-term $L_\infty$-algebra~\cite{Severa:2006aa}, see also~\cite{Jurco:2016qwv} for a detailed discussion of an extremely general case. Since we use these extensively, we present some more details in appendices~\ref{app:B},~\ref{app:C} and~\ref{app:D}.

Let us now come to the categorification of principal bundles. As explained in section~\ref{ssec:geometric_picture}, one can use any of the equivalent definitions of principal bundles and replace all notions by categorified ones. To describe transition functions of principal $\sG$-bundles, one can use the approach via functors from the {\em \v Cech groupoid} $\check \CCC(Y)$ of a surjective submersion $\sigma:Y\twoheadrightarrow M$ over some manifold $M$ to the category $\sB\sG=\sG\rightrightarrows *$. One may imagine $Y$ to be an open cover $Y=\sqcup U_i$, for concreteness sake. There are now two obvious projections from the fiber product $Y^{[2]}=Y\times_M Y$ (which is the space of double overlaps in the case $Y=\sqcup U_i$) to $Y$ as well as a diagonal embedding $Y\embd Y^{[2]}$, which, together with the composition $(y_1,y_2)\circ (y_2,y_3)=(y_1,y_3)$ we can use to form the 2-space
\begin{equation}
 \check \CCC(Y)=(Y^{[2]}\rightrightarrows Y)~.
\end{equation}
Since morphisms are invertible with $(y_1,y_2)^{-1}=(y_2,y_1)$, this is in fact a Lie groupoid. The transition functions of a principal $\sG$-bundle for some Lie group $\sG$ are given by functors $g:\check \CCC(Y)\rightarrow \sB\sG$ and isomorphism or gauge transformations correspond to natural transformations between functors. This picture readily extends to categorified groups, where we consider higher functors between the \v Cech groupoid, trivially regarded as a higher Lie groupoid, and the delooping $\sB\CCG$ of a higher Lie group $\CCG$. For $\CCG=\sB\sU(1)$, we recover abelian gerbes, or principal $\sB\sU(1)$-bundles in the form of {\em Hitchin--Chatterjee gerbes}~\cite{Chatterjee:1998}. These are stably isomorphic to Murray's more general and more useful {\em bundle gerbes}~\cite{Murray:9407015,Murray:2007ps}. Abelian gerbes, or principal $\sB\sU(1)$ bundles are described by a characteristic class, called the {\em Dixmier--Douady class} in $H^3(M,\RZ)$, which is the analogue of the first Chern class of line bundles in $H^2(M,\RZ)$.

To add connections, one can either glue together the local description as derived in section~\ref{ssec:hgt_dgas} or use insights from Lie differentiation, as done in~\cite{Jurco:2014mva,Jurco:2016qwv}. 

\subsection{\texorpdfstring{$L_\infty$}{Linfinity}-algebras and Lie \texorpdfstring{$n$}{n}-algebras}\label{app:B}

Strong homotopy Lie algebras or $L_\infty$-algebras comprise Lie algebras and are useful descriptions of all their categorifications. They play important roles in BV quantization, string field theory and higher geometry in general; the original references are~\cite{Zwiebach:1992ie,Lada:1992wc,Lada:1994mn}. 

An {\em $L_\infty$-algebra} is an $\RZ$-graded vector space $\sL=\oplus_{k\in\RZ}\sL_k$ which is endowed with a set of totally antisymmetric, multilinear maps $\mu_i:\wedge^i\sL\rightarrow\sL$, $i\in \NN$, of degree~$i-2$, which satisfy the {\em higher} or {\em homotopy Jacobi relations}
\begin{equation}\label{eq:homotopyJacobi}
 \sum_{r+s=i}\sum_{\sigma }(-1)^{rs}\chi(\sigma;\ell_1,\dots,\ell_{r+s})\mu_{s+1}(\mu_r(\ell_{\sigma(1)},\dots,\ell_{\sigma(r)}),\ell_{\sigma(r+1)},\dots,\ell_{\sigma(r+s)})\ =\ 0
\end{equation}
for all $\ell_1,\dots,\ell_{r+s}\in \sL$, where the second sum runs over all $(r,s)$ {\it unshuffles}, i.e.~permutations $\sigma$ of $\{1,\dots,r+s\}$ with the first $r$ and the last $s$ images of $\sigma$ ordered: $\sigma(1)<\dots<\sigma(r)$ and $\sigma(r+1)<\dots<\sigma(r+s)$. Also, $\chi(\sigma;\ell_1,\dots,\ell_i)$ denotes the {\it graded Koszul sign} defined through the graded antisymmetrized products
\begin{equation}
 \ell_1\wedge \dots \wedge \ell_i\ =\ \chi(\sigma;\ell_1,\dots,\ell_i)\,\ell_{\sigma(1)}\wedge \dots \wedge \ell_{\sigma(i)}~.
\end{equation}
If an $L_\infty$-algebra is nontrivial in degrees~$0,\dots,n-1$, we call it an $n$-term $L_\infty$-algebra. This is a useful and very general notion of a Lie $n$-algebra. We therefore often use the term Lie $n$-algebra, when we actually mean an $n$-term $L_\infty$-algebra. 

A {\em cyclic structure} on an $L_\infty$-algebra over $\FR$ is a graded symmetric, non-degenerate bilinear form 
\begin{equation}
 \langle -,-\rangle\,:\,\sL\odot \sL\ \rightarrow\ \FR~,
\end{equation}
such that 
\begin{equation}
 \langle\ell_1,\mu_i(\ell_2,\dots,\ell_{i+1})\rangle\ =\ (-1)^{i+|\ell_{i+1}|(|\ell_1|+\dots+|\ell_i|)}\langle\ell_{i+1},\mu_i(\ell_1,\dots,\ell_{i})\rangle
\end{equation}
for $\ell_1,\dots,\ell_{i+1}\in \sL$.

We are particularly interested in Lie 2- and 3-algebras. The lowest homotopy Jacobi relations are equivalent to the following ones:
\begin{equation}
\begin{aligned}
 \mu_1(\mu_1(\ell_1))&=0~,\\
 \mu_1(\mu_2(\ell_1,\ell_2))&=\mu_2(\mu_1(\ell_1),\ell_2)+(-1)^{|\ell_1|}\mu_2(\ell_1,\mu_1(\ell_2))~,\\
 \mu_1(\mu_3(\ell_1,\ell_2,\ell_3))&=-\mu_3(\mu_1(\ell_1),\ell_2,\ell_3)-(-1)^{|\ell_1|}\mu_3(\ell_1,\mu_1(\ell_2),\ell_3)-\\
 &~~~~-(-1)^{(|\ell_1|+|\ell_2|)}\mu_3(\ell_1,\ell_2,\mu_1(\ell_3))+\mu_2(\ell_1,\mu_2(\ell_2,\ell_3))-\\
 &~~~-\mu_2(\mu_2(\ell_1,\ell_2),\ell_3)-(-1)^{|\ell_1|\,|\ell_2|}\mu_2(\ell_2,\mu_2(\ell_1,\ell_3))~.
\end{aligned}
\end{equation}
The first two identities are the same as for a differential graded Lie algebra, but the third identity is a controlled lifting of the Jacobi identity. 

A cyclic structure on a Lie 3-algebra satisfies 
\begin{equation}
 \begin{aligned}
  \langle \ell_1,\mu_1(\ell_2)\rangle &= (-1)^{1+|\ell_2|\,|\ell_1|}\langle \ell_2,\mu_1(\ell_1)\rangle~,\\
  \langle \ell_1,\mu_2(\ell_2,\ell_3)\rangle &= (-1)^{1+|\ell_3|(|\ell_1|+|\ell_2|)+|\ell_1|\,|\ell_2|}\langle \ell_3,\mu_2(\ell_2,\ell_1)\rangle~,\\
  \langle \ell_1,\mu_3(\ell_2,\ell_3,\ell_4)\rangle &= (-1)^{1+|\ell_4|(|\ell_1|+|\ell_2|+\ell_3)+|\ell_3|(|\ell_1|+|\ell_2|)}\langle \ell_4,\mu_3(\ell_2,\ell_3,\ell_1)\rangle~,\\
  \langle \ell_1,\mu_4(\ell_2,\ell_3,\ell_4,\ell_5)\rangle &= (-1)^{1+|\ell_5|(|\ell_1|+|\ell_2|+\ell_3+\ell_4)+|\ell_4|(|\ell_1|+|\ell_2|+|\ell_3|)}\langle \ell_5,\mu_4(\ell_2,\ell_3,\ell_4,\ell_1)\rangle
 \end{aligned}
\end{equation}
for all $\ell_1,\dots,\ell_5\in \sL$.

\subsection{Lie \texorpdfstring{$n$}{n}-algebras as \texorpdfstring{N$Q$}{NQ}-manifolds}\label{app:C}

A very useful and elegant definition of $L_\infty$-algebras can be given in terms of N$Q$-manifolds, which are known to physicists from BRST quantization and string field theory. In this picture, the cyclic structure arises from a symplectic form. Below, we briefly explain this point of view.

An {\em N$Q$-manifold} $(\CM,Q)$ is an $\NN_0$-graded manifold $\CM$ endowed with a vector field $Q$ which is of degree~1 and nilquadratic: $Q^2=0$. Due to a similar argument as that for smooth (real) supermanifolds, N$Q$-manifolds can be regarded as $\NN$-graded vector bundles over the body $\CM_0$ of the manifold $\CM$. 

Archetypical examples of N$Q$-manifolds are $(T[1]M,\dd)$, the grade shifted tangent bundle together with the de~Rham differential as well as the grade shifted Lie algebra $(\frg[1],Q)$ with $Q=-\tfrac12 f_{\alpha\beta}^\gamma\xi^\alpha\xi^\beta \der{\xi^\gamma}$ a vector field of degree~1 in some coordinates $\xi^\alpha$ on $\frg[1]$, which are necessarily of degree~1. Note that $Q^2=0$ is equivalent to the Jacobi identity in the latter case.

The latter example indicates the relation of N$Q$-manifolds to Lie $n$-algebras: Given a Lie $n$-algebra $\sL$, we should first grade-shift the underlying graded vector space:
\begin{equation}
 \sL=(\sL_0\leftarrow \dots\leftarrow \sL_{n-1})~~~\rightarrow~~~\sL[1]=(*\leftarrow\sL_0[1]\leftarrow \dots\leftarrow \sL_{n-1}[1])~.
\end{equation}
Correspondingly, the degree of the maps $\mu_i$ changes from $i-2$ to $-1$. The sum of these maps forms a codifferential
\begin{equation}
 D=\mu_1+\mu_2+\mu_3+\dots~,
\end{equation}
which acts on the coalgebra $\wedge^\bullet \sL[1]$. If we now dualize to the algebra of functions on $\sL[1]$, the corresponding differential is a vector field of degree~$1$ on the N$Q$-manifold
\begin{equation}
 *\leftarrow\sL_0[1]\leftarrow \dots\leftarrow \sL_{n-1}[1]~.
\end{equation}
The condition $Q^2=0$ translates to~\eqref{eq:homotopyJacobi}.

Altogether, we arrived at a description of a Lie $n$-algebra $\sL$ as a {\em differential graded algebra} $\big(\CC^\infty(\sL[1]), Q\big)$. This differential graded algebra is also called the {\em Chevalley--Eilenberg algebra} ${\rm CE}(\sL)$ of $\sL$.

It is not hard to see that a cyclic structure on a Lie $n$-algebra $\sL$ corresponds to a symplectic form on $\sL[1]$ of degree~$n+1$, and we use this in section~\ref{ssec:ext_Lie_3}. For example, the inner product $(x_1,x_2)=g_{\alpha\beta}\xi_1^\alpha\xi_2^\beta$ on a Lie algebra $\frg$ is described by the symplectic form $\omega=\tfrac12 g_{\alpha\beta}\dd \xi^\alpha\wedge \dd \xi^\beta$ of degree~2 on $\frg[1]$.

\subsection{Morphisms and equivalences of Lie \texorpdfstring{$n$}{n}-algebras}\label{app:D}

Let us now come to categorical equivalence between Lie $n$-algebras. For simplicity, we first restrict ourselves to Lie 2-algebras $\sL=W\leftarrow V$. A morphism of Lie $n$-algebras, being a morphism of graded spaces, is most readily derived in the Chevalley--Eilenberg picture. There, a morphism $\Phi:\sL\rightarrow \tilde \sL$ between Lie 2-algebras $\sL=(W\leftarrow V)$ and $\tilde \sL=\tilde W\leftarrow \tilde V)$ is given by a morphism of differential graded algebras $\Phi^*$. That is, 
\begin{equation}
 \Phi^*:\CC^\infty(\tilde \sL[1])\rightarrow \CC^\infty(\sL[1])~,~~~\tilde Q\circ \Phi^*= \Phi^*\circ Q~.
\end{equation}
Since $\sL[1]$ is a graded vector space, this morphism is determined by its image on coordinate functions. Because degrees have to be preserved, $\Phi^*$ (and thus $\Phi$) is encoded in maps
\begin{equation}
 \phi_0:W\rightarrow \tilde W~,~~~\phi_0:V\rightarrow \tilde V~,~~~\phi_1:W\wedge W\rightarrow \tilde V~.
\end{equation}
The fact that $\Phi^*$ is a morphism of differential graded algebras implies that 
\begin{equation}\label{eq:L2_morphism}
\begin{aligned}
 \phi_0\left(\mu_2(w_1,w_2)\right)&=\tilde\mu_2\left(\phi_0(w_1),\phi_0(w_2)\right)+\tilde\mu_{1}(\phi_1(w_1,w_2))~,\\ 
\phi_0\left(\mu_2(w,v)\right)&=\tilde\mu_2(\phi_0(w),\phi_0(v))+\phi_1(w,\mu_{1}(v))~,\\
\phi_0\left(\mu_3(w_1,w_2,w_3)\right)&=\tilde\mu_3(\phi_0(w_1),\phi_0(w_2),\phi_0(w_3))+\left[
  \phi_1(w_1,\mu_2(w_2,w_3))\right. \\
&\hspace{2cm} \left.+\tilde\mu_2\left(\phi_0(w_1),\phi_1(w_2,w_3)\right)+\text{cyclic}\,(w_1,w_2,w_3)\right]
\end{aligned}
\end{equation}
for all $w,w_{1,2}\in W$ and $v\in V$. This reproduces the definition of a morphism of Lie 2-algebra from~\cite{Baez:2003aa}.

Two morphisms $\Phi=(\phi_0,\phi_1)$ and $\Psi=(\psi_0,\psi_1)$ compose as follows:
\begin{equation}\label{eq:comp_L2_morphisms}
 (\Psi\circ \Phi)_0(\ell)=\psi_0(\phi_0(\ell))~,~~~(\Psi\circ \Phi)_{1}(w_1,w_2)=\psi_0(\phi_1(w_1,w_2))+\psi_1(\phi_0(w_1),\phi_0(w_2))
\end{equation}
for all $\ell\in \sL$ and $w_{1,2}\in W$. The identity morphism reads as $\id_\sL=(\id_\sL,0)$. The above data of 2-term $L_\infty$-algebras and their morphisms, together with the identity morphism and composition combines to a category of Lie 2-algebras, $\CatLietwo$.

Note that the inverse of a morphism $\Phi=(\phi_0,\phi_1)$ is defined if and only if $\phi_0$ is invertible:
\begin{equation}
 (\Phi^{-1})_0(\ell)=\phi_0^{-1}(\ell)~,~~~(\Phi^{-1})_{-1}(w_1,w_2)=-\phi_0^{-1}(\phi_1(\phi_0^{-1}(w_1),\phi_0^{-1}(w_2)))~,
\end{equation}
again for $\ell\in \sL$ and $w_{1,2}\in W$.

A {\em 2-morphism} between two morphisms $\Phi,\Psi:\sL\rightarrow \tilde \sL$ is a chain homotopy $\chi:W\rightarrow \tilde V$ such that
\begin{equation}
 \phi_1(w_1,w_2)-\psi_1(w_1,w_2)=\mu_2(w_1,\chi(w_2))+\mu_2(\chi(w_1),\psi_0(w_2))-\chi(\mu_2(w_1,w_2))
\end{equation}
for all $w_1,w_2$.

Finally, an {\em equivalence} or a {\em quasi-isomorphism of Lie $2$-algebras} between Lie 2-algebras $\sL=W\xleftarrow{~\mu_1~}V$ and $\tilde\sL=\tilde W\xleftarrow{~\tilde\mu_1~}\tilde V$ is a morphism $\Phi_1:\sL\rightarrow \tilde \sL$ and a morphism $\Phi_2:\tilde \sL\rightarrow \sL$ such that $\Phi_1\circ \Phi_2\cong \unit_{\tilde frg}$ and $\Phi_2\circ \Phi_1\cong \unit_{\sL}$. An explicit example of an equivalence between Lie 2-algebras is found in section~\ref{ssec:string_Lie_2_algebra}. 

The above discussion readily generalizes to morphisms $\Phi$ between Lie $n$-algebras $\sL$ and $\tilde \sL$. These are given by totally antisymmetric maps $\phi_i:\sL^{\wedge i+1}\rightarrow \sL$, $i=0,\dots,n-1$, of degree~$-i$ such that appropriate extensions of~\eqref{eq:L2_morphism} hold, see~\cite{Fregier:2013dda} for details. An {\em isomorphism of Lie $n$-algebras} is a morphism of Lie $n$-algebras with $\phi_0$ an isomorphisms. Equivalences between $L_\infty$-algebras are then captured by {\em Lie $n$-algebra quasi-isomorphisms} which are morphisms of Lie $n$-algebras which induce an isomorphism on the cohomology of the complex underlying the Lie $n$-algebras.

\bibliography{bigone}

\bibliographystyle{latexeu}

\end{document}